\documentclass[a4paper,onecolumn,11pt,accepted=2020-10-12]{quantumarticle}
\pdfoutput=1

\usepackage[utf8]{inputenc}
\usepackage[english]{babel}
\usepackage[T1]{fontenc}
\usepackage{amsmath,amsfonts}
\usepackage{hyperref}
\usepackage{graphicx}
\usepackage{braket}
\usepackage{hyperref}

\newcommand{\mtx}[1]{\boldsymbol{\mathsf{#1}}}
\renewcommand{\vec}[1]{\boldsymbol{#1}}
\newcommand{\unit}[1]{\hat{\boldsymbol{#1}}}

\begin{document}

\title{Emergence of the Born rule in quantum optics}

\author{Brian R. La Cour}
\affiliation{Applied Research Laboratories, The University of Texas at Austin, P.O. Box 8029, Austin, TX 78713-8029}
\email{blacour@arlut.utexas.edu}
\author{Morgan C. Williamson}

\maketitle

\begin{abstract}
The Born rule provides a fundamental connection between theory and observation in quantum mechanics, yet its origin remains a mystery.  We consider this problem within the context of quantum optics using only classical physics and the assumption of a quantum electrodynamic vacuum that is real rather than virtual.  The connection to observation is made via classical intensity threshold detectors that are used as a simple, deterministic model of photon detection.  By following standard experimental conventions of data analysis on discrete detection events, we show that this model is capable of reproducing several observed phenomena thought to be uniquely quantum in nature, thus providing greater elucidation of the quantum-classical boundary.
\end{abstract}


\section{Introduction}

Since the appearance of Bell's inequality, it has become apparent that local hidden variable models cannot be compatible with the complete mathematical formalism of quantum mechanics \cite{vonNeumann1931,Bell1964,deMuynck2002,Brunner2014}.  Indeed, recent loophole-free experiments appear to be consistent with this conclusion \cite{Hensen2015,Giustina2015,Shalm2015,Rosenfeld2017}.  Nevertheless, there remains the open question of which observed phenomena, in particular, are truly quantum in nature and have no classical analogue.  This question of elucidating the quantum-classical boundary is of practical importance, as many new and emerging technologies, such as quantum computing, quantum communication, and quantum sensing, rely upon this distinction for their efficacy and security \cite{Bierhorst2018}.

The field of quantum optics would seem to be a good place to explore this question, as the systems of interest are relatively simple to describe in terms of discrete field modes, while the important light-matter interactions may be restricted to the physics of photodetection devices.  One of the more curious aspects of quantum optics is the concept of the vacuum or zero-point field (ZPF).  In quantum electrodynamics (QED), a vacuum state is defined simply to be the lowest energy state of a given field mode \cite{Landau&Lifshitz4}.  The number of photons in this state is taken to be zero, yet its energy is nonzero, giving rise to the notion of ``virtual'' photons.  Although the quantum vacuum is viewed as being only virtual, its effects are quite real.  Phenomena such as the Casimir force, van der Waals attraction, Lamb shifts, and spontaneous emission are all believed to have their origin in the quantum vacuum \cite{Milonni1994}.

The prominence of vacuum states in quantum optics suggests that they may be useful in developing a physical theory that explores the quantum-classical boundary.  In this work, we shall proceed by supposing that the quantum vacuum of QED is real, not virtual.  In doing so, we shall abandon all formal reference to quantum theory and consider a world governed solely by classical physics, albeit one in which the presence of a reified vacuum field is unavoidable.  Our connection to quantum theory will lie solely in the demand that the statistical description of the real vacuum field match that of the virtual one.  Our goal in doing so will be to explore which observed quantum phenomena can be explained under this supposition.  In particular, we shall explore in this work the emergence of the Born rule as a statistical prediction that is applicable only within a certain regime of validity and application.

Several previous attempts have been made to derive the Born rule from first principles \cite{Landsman2008}.  Max Born, in his original 1926 paper, considered the problem of perturbative scattering and suggested that the resulting energy may be interpreted as a statistical average if the scattering amplitudes, when properly squared, are interpreted as probabilities \cite{Born1926}.  Gleason provided the first attempt at a mathematical derivation of the Born rule but relied on an assumed association of Hermitian operators with measurement observables \cite{Gleason1957}.  David Deutsch, in 1999, went further to argue that elementary decision theory may be used to deduce the Born rule as a necessary consequence of the other quantum axioms \cite{Deutsch1999}.  This argument has since been criticized to be circular, as it requires the assumption of an agent with a particular predilection for $L_2$ norms \cite{Barnum2000}.  Zurek has suggested decoherence as an explanation of the Born rule \cite{Zurek2005}, although this view has been criticized as well to be insufficient \cite{Schlosshauer2005}.  More recently, Masanes \textit{et al.} have claimed to derive the Born rule by assuming, among other things, that measurements consist of well-defined trials and always produce one of a pre-defined set of outcomes \cite{Masanes2019}.  While seemingly innocuous, this assumption does not always hold in real, experimental settings where, for example, photons are detected at random times or, more often, not at all.  An interesting result from Allahverdyan \textit{et al.} provides a derivation of the Born rule from the dynamical law of quantum mechanics with the context of spin systems \cite{Allahverdyan2013}.  

Working within the confines of the formalism does not seem a promising approach to deriving physical laws.  What these and other attempts to derive the Born rule lack is any attempt to model the actual physics of measurement.  This paper seeks to address that point by considering a deterministic model of measurement together with a reified quantum vacuum.

A reified quantum vacuum is the premise behind the theory of stochastic electrodynamics (SED), and we adopt a similar outlook here \cite{TheDice}.  Previous work in SED considered the statistical behavior of physical systems immersed in the zero-point field.  These included classical descriptions of the quantum harmonic oscillator ground state as well as spontaneous parametric downconversion \cite{Casado1997}.  Although these efforts were successful insofar as they predicted probability density functions identical to the corresponding quantum Wigner function, they failed to fully appreciate the critical role of measurement and experimental procedure in the observation of quantum phenomena.  In particular, the role of post-selection and its relation to contextuality has received little attention within SED.

To address this deficiency, we shall consider here a local, deterministic model of photon detection wherein the only random variables determining the outcome of a measurement are those associated with the relevant vacuum states incident upon the device.  This approach differs from previous work in stochastic optics, an offshoot of SED focused on quantum optics, wherein the intensity of incident waves above a given threshold determines only the \emph{probability} of an outcome, leaving the actual realization to be determined by yet another, implicit, hidden variable \cite{Marshall1988}.  Our approach uses a deterministic amplitude threshold crossing scheme to define detection events and is similar to the work of other researchers in this regard \cite{Adenier2009, LaCour2014, Khrennikov}.  A key difference from previous work is the use of post-selection and the examination of asymptotic behavior to approximate ideal quantum predictions.

The structure of the paper is as follows.  In section \ref{sec:RVF} we describe the mathematical model used to describe the reified vacuum field and use the single-mode approximation to make the correct correspondence with quantum optics.  The connection to observation is made in section \ref{sec:ATD}, where we describe a deterministic model of quantum measurement using amplitude threshold detection.  From this, the Born rule is shown to arise as an emergent and approximate property of the model in the presence of measurements.  Finally, in section \ref{sec:GLT} we consider the general problem of transformations of multiple vacuum modes under linear optics to arrive at a model approximating single-photon, multi-mode quantum states.  Conclusions are summarized in section \ref{sec:C}.


\section{The Reified Vacuum Field}
\label{sec:RVF}


\subsection{Continuum Description}

Any classical electric field may be written in terms of a continuum of plane wave modes.  Thus, the electric field at a point $\vec{x}$ and time $t$ may be written, in Gaussian units, as
\begin{equation}
\vec{E}(\vec{x},t) = \frac{1}{2\pi} \int \sqrt{\mathcal{E}(\vec{k})} \, \vec{a}(\vec{k}) \, e^{i\vec{k}\cdot\vec{x}-i\omega(\vec{k}) t} \, d\vec{k} + \mathrm{c.c.} \; ,
\label{eqn:Econt}
\end{equation}
where $\mathcal{E}(\vec{k}) \ge 0$ is a scale factor related to the modal energy for wave vector $\vec{k} \in \mathbb{R}^3$, $\vec{a}(\vec{k}) \in \mathbb{C}^3$ gives the field direction and phase, and $\omega(\vec{k}) \ge 0$ is the angular frequency.  For a classical vacuum, $\omega(\vec{k}) = \|\vec{k}\|c$, where $\|\vec{k}\|$ is the magnitude of $\vec{k}$ and $c$ is the speed of light.  The term ``c.c.'' represents the complex conjugate of the term to the left.  The magnetic field is similarly described, with $\vec{a}(\vec{k})$ replaced by $\vec{k} \times \vec{a}(\vec{k})$, so that specifying $\mathcal{E}(\vec{k})$, $\vec{a}(\vec{k})$, and $\omega(\vec{k})$ for all $\vec{k} \in \mathbb{R}^3$ provides a complete description of the electromagnetic field.  Without loss of generality, we shall take $\vec{a}(\vec{k})$ to be stochastic, while $\mathcal{E}(\vec{k})$ is assumed fixed.

For convenience, we may decompose $\vec{a}(\vec{k})$ into orthogonal polarization modes.  For each wave vector $\vec{k}$, let $\unit{e}_0(\vec{k})$ and $\unit{e}_1(\vec{k})$ be any two orthogonal polarization vectors (i.e., complex vectors such that $\unit{k} \cdot \unit{e}_{\mu}(\vec{k}) = 0$, where $\unit{k} = \vec{k}/\|\vec{k}\|$, and $\unit{e}_{\mu}(\vec{k})^* \cdot \unit{e}_{\nu}(\vec{k}) = \delta_{\mu,\nu}$ for all $\mu, \nu \in \{0,1\}$).  We may then write
\begin{equation}
\vec{a}(\vec{k}) = a_0(\vec{k}) \, \unit{e}_0(\vec{k}) + a_1(\vec{k}) \, \unit{e}_1(\vec{k}) \; ,
\end{equation}
where $a_{\mu}(\vec{k}) = \unit{e}_{\mu}(\vec{k})^* \cdot \vec{a}(\vec{k}) \in \mathbb{C}$.  Note that the choice of polarization vectors is arbitrary, and may vary with $\vec{k}$, but is otherwise taken to be fixed and nonrandom.

We now turn to the correspondence with quantum theory.  Consistency with quantum electrodynamics will require that, at zero temperature,
\begin{equation}
\mathcal{E}(\vec{k}) = \mathcal{E}_0(\omega(\vec{k})) := \tfrac{1}{2}\hbar\omega(\vec{k}) \; ,
\end{equation}
where we have now introduced $\hbar$, Planck's constant divided by $2\pi$, as setting the fundamental scale of the vacuum field.  For nonzero temperatures, $\mathcal{E}_0(\omega)$ is replaced by the expression
\begin{equation}
\mathcal{E}_T(\omega) = \tfrac{1}{2} \hbar\omega + \frac{\hbar\omega}{e^{\hbar\omega/k_BT}-1} \; , 
\label{eqn:PlanckT}
\end{equation}
where $\omega > 0$, $k_B$ is Boltzmann's constant, and $T > 0$ is the absolute temperature.  Since the density of states is given by $\omega^2/(\pi^2 c^3)$, the spectral energy density is
\begin{equation}
\rho_T(\omega) = \frac{\hbar \omega^3}{\pi^2 c^3} \left( \tfrac{1}{2} + \frac{1}{e^{\hbar\omega/k_BT}-1} \right) \; ,
\end{equation}
which corresponds to Planck's ``second quantum theory'' of blackbody radiation, with a zero-point energy term included \cite{Planck1911}.  Note also that, at zero temperature, $\rho_0(\omega) = \hbar\omega^3/(2\pi^2c^3)$ is Lorentz invariant, owing to the cubic dependence on frequency, so the spectral energy density is the same in all inertial reference frames \cite{Marshall1963}.

The stochastic nature of the field is described entirely by $\vec{a}(\cdot)$, and consistency with QED requires that it be a complex Gaussian random vector field such that, for any choice of polarization vectors, $\mathsf{E}[a_{\mu}(\vec{k})] = 0$ and
\begin{subequations}
\begin{align}
\mathsf{E}[a_{\mu}(\vec{k}) \, a_{\nu}(\vec{k}')^*] &= \delta_{\mu,\nu} \, \delta(\vec{k}-\vec{k}') \\
\mathsf{E}[a_{\mu}(\vec{k}) \, a_{\nu}(\vec{k}')] &= 0 \; ,
\end{align}
\end{subequations}
where $\mathsf{E}[\cdot]$ denotes an expectation value \cite{Ibison1996}.  More generally, the $n$-point correlations of the field are given by
\begin{equation}
\mathsf{E}\left[ \prod_{i=1}^{n} a_{\mu_i}(\vec{k}_i) \, a_{\nu_i}(\vec{k}_i')^* \right] = 
\prod_{i=1}^{n} \prod_{j=i}^{n} \delta_{\mu_i,\nu_j} \, \delta(\vec{k}_i-\vec{k}_j') \; ,
\end{equation}
with all other combinations giving a zero expectation value.  Of course, this mathematical correlation structure is only an idealization; on \emph{some} spatio-temporal scale, the field must surely be correlated.  We would furthermore expect that the statistical character of the field, its scale and correlations, might also change over time and space.  Nevertheless, we shall proceed with this modest idealization of the zero-point field, as it will provide a useful model for the quantum vacuum.


\subsection{Discrete-Mode Approximation}

One can approximate the continuum of wave vector modes by a set of closely spaced discrete modes in a notional box.  Given a cube of length $L > 0$, we define a set $K$ of discrete-mode wave vectors as follows:
\begin{equation}
K = \left\{ \frac{2\pi n_1}{L}\unit{x} + \frac{2\pi n_2}{L}\unit{y} + \frac{2\pi n_3}{L}\unit{z} \; : \; n_1, n_2, n_3 \in \mathbb{Z} \right\} \; .
\end{equation}
The continuum wave vector space may now be decomposed into notional discrete cells  $C(\vec{k}) = \vec{k} + [-\frac{\pi}{L}, \frac{\pi}{L})^3$, each centered on a wave vector $\vec{k} \in K$.  Since the cells are disjoint and their union comprises all of $\mathbb{R}^3$, we may rewrite equation (\ref{eqn:Econt}) as follows:
\begin{equation}
\!\!\!\!\! \!\!\!\!\! \!\!\!\!\!
\vec{E}(\vec{x},t) = \frac{1}{2\pi} \sum_{\vec{k} \in K} \sum_{\mu} \int_{C(\vec{k})} \sqrt{\mathcal{E}_0(\omega(\vec{k}'))} \, a_{\mu}(\vec{k}') \, \unit{e}_{\mu}(\vec{k}') \, e^{i\vec{k}'\cdot\vec{x}-i\omega(\vec{k}')t} \, d\vec{k}' + \mathrm{c.c.}
\end{equation}
Furthermore, if the cells are small (i.e., $L$ is large), we may make the approximation
\begin{equation}
\!\!\!\!\! \!\!\!\!\!
\vec{E}(\vec{x},t) \approx \frac{1}{2\pi} \sum_{\vec{k} \in K} \sum_{\mu} \sqrt{\mathcal{E}_0(\omega(\vec{k}))} \, \unit{e}_{\mu}(\vec{k})\, e^{i\vec{k}\cdot\vec{x}-i\omega(\vec{k})t} \int_{C(\vec{k})} a_{\mu}(\vec{k}') \, d\vec{k}' + \mathrm{c.c.}
\end{equation}

This last integral yields a complex Gaussian random variable with zero mean and a variance of $\Delta\vec{k} = (2\pi/L)^3 = 8\pi^3/V$ corresponding to the volume of each cell \cite{Lasota&Mackey}.  We may therefore write
\begin{equation}
\int_{C(\vec{k})} a_{\mu}(\vec{k}') \, d\vec{k}' = z_{\mu,\vec{k}} \sqrt{\Delta\vec{k}} \; ,
\end{equation}
where $z_{\mu,\vec{k}}$ is a standard complex Gaussian random variable (i.e., a complex Gaussian random variable such that $\mathsf{E}[z_{\mu,\vec{k}}] = 0$, $\mathsf{E}[|z_{\mu,\vec{k}}|^2] = 1$, and $\mathsf{E}[z_{\mu,\vec{k}}^2] = 0$).    Equivalently, we may write $z_{\mu,\vec{k}}$ in the form $z_{\mu,\vec{k}} = (x + i y)/\sqrt{2}$, where $x, y$ are independent, real-valued standard normal random variables.

We note that all discrete modes differing in either wave vector or polarization are independent since, for $\vec{k}, \vec{k}' \in K$,
\begin{equation}
\begin{split}
\mathsf{E}[z_{\mu,\vec{k}} \, z_{\nu,\vec{k'}}^*] &= \frac{1}{\Delta\vec{k}} \int_{C(\vec{k})} \int_{C(\vec{k}')} \mathsf{E}[a_{\mu}(\vec{k}'')^* a_{\nu}(\vec{k}''')] \, d\vec{k}'' d\vec{k}''' \\
&= \frac{1}{\Delta\vec{k}} \int_{C(\vec{k})} \int_{C(\vec{k}')} \delta_{\mu,\nu} \, \delta(\vec{k}'''-\vec{k}'') \, d\vec{k}'' d\vec{k}''' \\
&= \delta_{\mu,\nu} \, \frac{1}{\Delta\vec{k}} \int_{C(\vec{k}) \cap C(\vec{k}')} d\vec{k}'' \\
&= \delta_{\mu,\nu} \, \delta_{\vec{k},\vec{k}'} \; .
\end{split}
\end{equation}

We shall chiefly be concerned with descriptions in terms of the discrete-mode approximation, as this affords the clearest correspondence with quantum optics.  In particular, the lowering operator $\hat{a}_{\mu,\vec{k}}$ for discrete mode $(\mu,\vec{k})$ may be associated with the random variable $z_{\mu,\vec{k}}/\sqrt{2}$ in the sense that the vacuum expectation of the symmetrized number operator equals the variance of the corresponding random variable.  To see this, observe that
\begin{equation}
\bra{0} \tfrac{1}{2} \Bigl( \hat{a}_{\mu,\vec{k}}^\dagger \hat{a}_{\mu,\vec{k}} + \hat{a}_{\mu,\vec{k}} \hat{a}_{\mu,\vec{k}}^\dagger \Bigr) \ket{0} = \tfrac{1}{2} (0 + 1) = \tfrac{1}{2} \; ,
\end{equation}
and, similarly,
\begin{equation}
\mathsf{E}\left[ \tfrac{1}{2} \left( \frac{z_{\mu,\vec{k}}^*}{\sqrt{2}} \frac{z_{\mu,\vec{k}}}{\sqrt{2}} + \frac{z_{\mu,\vec{k}}}{\sqrt{2}} \frac{z_{\mu,\vec{k}}^*}{\sqrt{2}} \right) \right] = \mathsf{E}\left[ \left|\frac{z_{\mu}(\vec{k})}{\sqrt{2}}\right|^2 \right] = \tfrac{1}{2} \; .
\end{equation}
Note that, since $z_{\mu,\vec{k}}$ and $z_{\mu,\vec{k}}^*$ commute, whereas $\hat{a}_{\mu,\vec{k}}$ and $\hat{a}_{\mu,\vec{k}}^\dagger$ do not, symmetrization of the operators is important to achieve the correct correspondence.  

The connection to quantum optics can be further elucidated by examining the modal energy.  Quantum mechanically, the energy of the vacuum state is given by the expectation value of the Hamilitonian $\hat{H} = \frac{1}{2} ( \hat{a}_{\mu,\vec{k}}^\dagger \hat{a}_{\mu,\vec{k}} + \hat{a}_{\mu,\vec{k}} \hat{a}_{\mu,\vec{k}}^\dagger ) \, \hbar\omega(\vec{k})$, which is simply the symmetrized number operator scaled by $\hbar\omega(\vec{k})$.  Thus, $\bra{0} \hat{H} \ket{0} = \frac{1}{2} \hbar\omega(\vec{k})$ is identified as the average energy per vacuum mode.

To find the corresponding classical value, we begin by computing the energy density of the electromagnetic field for the selected mode, as given by
\begin{equation}
u(\vec{x},t) = \frac{1}{8\pi} \Bigl[ \|\Delta\vec{E}(\vec{x},t)\|^2 + \|\Delta\vec{B}(\vec{x},t)\|^2 \Bigr] \; ,
\end{equation}
where the single-mode electric field is
\begin{equation}
\Delta\vec{E}(\vec{x},t) = \frac{1}{2\pi} \sqrt{\mathcal{E}_0(\omega)} \, z_{\mu,\vec{k}} \, \unit{e}_{\mu}(\vec{k}) \, e^{i\vec{k} \cdot \vec{x} - i\omega t} \, \sqrt{\Delta\vec{k}} + \mathrm{c.c.}
\end{equation}
and the single-mode magnetic field is
\begin{equation}
\Delta\vec{B}(\vec{x},t) = \frac{1}{2\pi} \sqrt{\mathcal{E}_0(\omega)} \, z_{\mu,\vec{k}} \, [\unit{k} \times \unit{e}_{\mu}(\vec{k})] \, e^{i\vec{k} \cdot \vec{x} - i\omega t} \, \sqrt{\Delta\vec{k}} + \mathrm{c.c.} \; .
\end{equation}
Using the fact that, for any complex vector $\vec{v}$, $\|\vec{v}+\vec{v}^* \|^2 = 2( \vec{v}^*\cdot\vec{v} + \mathrm{Re}[\vec{v}\cdot\vec{v}])$, we find that
\begin{equation}
\|\Delta\vec{E}(\vec{x},t)\|^2 = \frac{2\pi\hbar\omega(\vec{k})}{V} \, \left( |z_{\mu,\vec{k}}|^2  + \mathrm{Re}\left[ z_{\mu,\vec{k}}^2 e^{i2(\vec{k}\cdot\vec{x} - \omega(\vec{k}) t)} \right] \right) \; ,
\end{equation}
and, since $\|\Delta\vec{E}(\vec{x},t)\|^2 = \|\Delta\vec{B}(\vec{x},t)\|^2$, we conclude that
\begin{equation}
u(\vec{x},t) = \frac{\hbar\omega(\vec{k})}{2V} \, \left( |z_{\mu,\vec{k}}|^2  + \mathrm{Re}\left[ z_{\mu,\vec{k}}^2 e^{i2(\vec{k}\cdot\vec{x} - \omega(\vec{k}) t)} \right] \right)  \; .
\end{equation}

Now consider the time average of $u(\vec{x},t)$, a spatially independent random variable given by
\begin{equation}
\bar{u} = \frac{\omega(\vec{k})}{2\pi} \int_0^{2\pi/\omega(\vec{k})} u(\vec{x},t) \, dt = \frac{\hbar\omega(\vec{k})}{2V} \, |z_{\mu,\vec{k}}|^2 \; .
\label{eqn:Bob}
\end{equation}
The expectation value of this time average over realizations of the ZPF is therefore
\begin{equation}
\mathsf{E}[\bar{u}] = \tfrac{1}{2} \hbar\omega(\vec{k}) / V \; .
\end{equation}

This result matches the quantum mechanical prediction if one integrates over a box of volume $V$ to find the total expected energy.  Of course, this volume is only notional and arises as an artifact of our discrete-mode approximation.  It describes the degree to which the single-mode approximation is valid rather than any physical volume.  For, say, a conical beam with a small half-angle of $\Delta\theta$ and a filtered bandwidth of $\Delta\omega$, we have $\Delta\vec{k} = \pi \Delta\theta^2 \Delta\omega/c$.  Thus, as the beam is narrowed, the notional volume increases and the energy density decreases proportionally.  An equivalent, and perhaps more physically meaningful, interpretation of the quantum mechanical energy, then, might be that the quantity $\bra{0} \hat{H} \ket{0}/V$ gives the expected energy density of a single vacuum mode when the wave vector is filtered and collimated to a resolution of $\Delta\vec{k} = 8\pi^3/V$.  In this context, we may there identify a correspondence between the Hamiltonian operator $\hat{H}$ and the time-averaged classical electromagnetic energy $H = \bar{u} V$.


\subsection{Coherent States}
\label{ssec:Corey}

In quantum optics, coherent states are considered the closest analogue to a classical state.  Previous work in SED has identified coherent states as arising from, for example, classical driven harmonic oscillators coupled to the ZPF \cite{Franca&Marshall1988}.  Here we shall consider an optical analogue in which we add a classical plane wave to a single mode of the ZPF.  

Recall that, previously, we had defined the ZPF to be of the form
\begin{equation}
\vec{E}(\vec{x},t) = \frac{1}{2\pi} \int \sqrt{\mathcal{E}_0(\omega(\vec{k}))} \, \vec{a}(\vec{k}) \, e^{i\vec{k}\cdot\vec{x}-i\omega(\vec{k}) t} \, d\vec{k} + \mathrm{c.c.} \; .
\end{equation}
We now add to this a plane wave with wave vector $\vec{k}_0$ and polarization $\unit{e}_0$ of the form
\begin{equation}
\vec{F}(\vec{x},t) = E_0 \, \unit{e}_0 \, e^{i\vec{k}_0\cdot\vec{x} - i\omega_0 t} + E_0^* \, \unit{e}_0^* \, e^{-i\vec{k}_0\cdot\vec{x} - i\omega_0 t} \; ,
\end{equation}
where $E_0 \in \mathbb{C}$ is a complex number representing the amplitude and phase of the external plane wave.  The total electric field is now
\begin{equation}
\vec{F}(\vec{x},t) + \vec{E}(\vec{x},t) \\
=  \int \left[ E_0 \, \unit{e}_0 \, \delta(\vec{k} - \vec{k}_0) + \frac{1}{2\pi} \sqrt{\mathcal{E}_0(\omega(\vec{k}))} \, \vec{a}(\vec{k}) \right] e^{i(\vec{k}\cdot\vec{x}-\omega t)} \, d\vec{k} \ + \mathrm{c.c.}
\end{equation}

In the single-mode approximation with $\Delta\vec{k} = 8\pi^3/V$, the total field becomes
\begin{equation}
\vec{F}(\vec{x},t) + \Delta\vec{E}(\vec{x},t) \approx \left[ E_0 + \frac{1}{2\pi} \sqrt{\mathcal{E}_0(\omega_0)} \, z \, \sqrt{\Delta\vec{k}} \right] \unit{e}_0 \, e^{i(\vec{k}_0\cdot\vec{x} - \omega_0 t)} + \mathrm{c.c.} \; 
\end{equation}
where $z$ is a standard complex Gaussian random variable.  For reasons that will soon become apparent, we shall express $E_0$ in the form
\begin{equation}
E_0 = \alpha\sqrt{\frac{2\pi\hbar\omega_0}{V}} \; ,
\end{equation}
where $\alpha \in \mathbb{C}$ is a complex number that will later be identified as the coherent state parameter.  The combined field in the single-mode approximation may now be written
\begin{equation}
\vec{F}(\vec{x},t) + \Delta\vec{E}(\vec{x},t) = \sqrt{\frac{2\pi\hbar\omega_0}{V}} \left( \alpha + \frac{z}{\sqrt{2}} \right) \unit{e}_0 \, e^{i(\vec{k}_0\cdot\vec{x} - \omega_0 t)} + \mathrm{c.c.}
\end{equation}

The energy density of the corresponding electromagnetic field is
\begin{equation}
\begin{split}
u(\vec{x},t) &= \frac{1}{8\pi} \|\vec{F}(\vec{x},t) + \Delta\vec{E}(\vec{x},t)\|^2 \times 2 \\
&= \frac{1}{4\pi} \left[ \|\vec{F}(\vec{x},t)\|^2 + \|\Delta\vec{E}(\vec{x},t)\|^2 + 2 \vec{F}(\vec{x},t) \cdot \Delta\vec{E}(\vec{x},t) \right] \; ,
\end{split}
\end{equation}
where
\begin{equation}
\|\vec{F}(\vec{x},t)\|^2 = \frac{4\pi\hbar\omega_0}{V} \left( |\alpha|^2 + \mathrm{Re}\left[ \alpha^2 e^{i2(\vec{k}_0\cdot\vec{x} - \omega_0 t)} \right] \right)
\end{equation}
\begin{equation}
\|\Delta\vec{E}(\vec{x},t)\|^2 = \frac{4\pi\hbar\omega_0}{V} \, \frac{1}{2} \left( |z|^2  + \mathrm{Re}\left[ z^2 e^{i2(\vec{k}_0\cdot\vec{x} - \omega_0 t)} \right] \right)
\end{equation}
\begin{equation}
\vec{F} \cdot \Delta\vec{E}(\vec{x},t) = \frac{4\pi\hbar\omega_0}{V} \, \frac{1}{\sqrt{2}} \mathrm{Re}\left[ \alpha z^* + \alpha z \, e^{i2(\vec{k}_0\cdot\vec{x} - \omega_0 t)} \right] \; .
\end{equation}
Taking the time average of $u(\vec{x},t)$ gives
\begin{equation}
\bar{u} = \frac{\hbar\omega_0}{V} \left( |\alpha|^2 + |z|^2 + \sqrt{2} \mathrm{Re}[\alpha z^*] \right) = \left| \alpha + \frac{z}{\sqrt{2}} \right|^2 \frac{\hbar\omega_0}{V} \; ,
\label{eqn:Charlie}
\end{equation}
and the expectation value of $\bar{u}$ over realizations of the ZPF is
\begin{equation}
\mathsf{E}[\bar{u}] = \left( |\alpha|^2 + \tfrac{1}{2} \right) \frac{\hbar\omega_0}{V} \; .
\end{equation}
This result matches the familiar energy density $\bra{\alpha} \hat{H} \ket{\alpha}/V$ of a quantum optical coherent state $\ket{\alpha}$.

For general thermal states, $\mathcal{E}_0$ is replaced by $\mathcal{E}_T$, as defined in equation (\ref{eqn:PlanckT}), and, hence, $z/\sqrt{2}$ is replaced by the scaled quantity $\sigma z$, where
\begin{equation}
\sigma = \sqrt{ \mathcal{E}_T(\omega_0)/(\hbar\omega_0) } \; .
\end{equation}
In this case, the average energy density becomes $(|\alpha|^2 + \sigma^2) \hbar\omega_0/V$.  Note that nonzero temperatures merely have the effect of rescaling the ZPF for the given mode.  At high temperatures ($\sigma \gg |\alpha|$), the coherent state becomes indistinguishable from thermal noise.  Conversely, at large amplitudes ($|\alpha| \gg \sigma$), the coherent state becomes indistinguishable from a classical plane wave of fixed amplitude and phase.


\section{Amplitude Threshold Detection}
\label{sec:ATD}

We have described a mathematical model for the QED vacuum in terms of a  stochastic electromagnetic field.  To make the important connection to observation and discrete detection events, we now introduce a simple deterministic model of photon detection based on amplitude threshold crossings and motivated by the observed behavior of real detectors.  

Suppose that we have, to arbitrary precision, isolated a single angular frequency $\omega_0$, polarization mode $\unit{e}_0$, and wave vector mode $\vec{k}_0$ of the vacuum in the discrete-mode approximation with wave vector resolution $\Delta\vec{k}$.  For the vacuum and coherent states, the energy density $u(\vec{x},t)$ at position $\vec{x}$ and time $t$ varies sinusoidally in time and space.  We imagine a detection device that reacts slowly enough as to be sensitive only to the time average, $\bar{u}$, of the energy density and note that this averaging eliminates both the temporal and spatial dependence of the energy density.   Using a time average is justified by the fact that a typical period of light is orders of magnitude shorter than the corresponding lag time for the photoelectric effect \cite{Ossiander2018}.

Now, although $\bar{u}$ is constant across space and time, it varies from one vacuum field realization to another due to the presence of the random variable $z$.  We now ask whether this time-averaged energy density falls above some threshold $\Gamma^2 \ge 0$.  Such an outcome will be deemed a detection event or ``click'' of a detector, and the probability of such an event occurring will be denoted $\Pr[\bar{u} > \Gamma^2]$.  Note that the vacuum realization $z$ is the only source of randomness in determining this probability, and, in the single-mode limit, the coherence time of the given vacuum mode is infinite.


\subsection{Dark Counts}

The time-averaged energy density of the vacuum is given by equation (\ref{eqn:Charlie}) with $\alpha = 0$ and, being the sum of two squared independent normal distributions, has an exponential distribution with a mean of $\frac{1}{2}\hbar\omega_0/V$.  The probability of a detection event is therefore
\begin{equation}
\Pr\left[ \bar{u} > \Gamma^2 \right] = \exp(-2V\Gamma^2/\hbar\omega_0) \; .
\end{equation}

Since we have assumed $V$ is large, we may take $\Gamma^2$ to be comparably small.  In particular, we shall adopt the \emph{single-mode limit}, analogous to the thermodynamic limit, in which
\begin{equation}
\lim_{V \to \infty} \frac{V\Gamma^2}{\hbar\omega} = \gamma^2
\end{equation}
for some $\gamma \ge 0$.  (Note that $\gamma$ may be specific to a particular polarization, frequency and wave vector resolution.)  In the single-mode limit, the probability of a detection event is $\exp(-2\gamma^2)$, which we interpret as the probability of a dark count for the vacuum state at zero temperature.  In a thermal state ($T > 0$), we replace $\frac{1}{2}\hbar\omega$ with $\sigma^2\hbar\omega$, so the probability of a dark count becomes $\exp(-\gamma^2/\sigma^2)$.  This, again, becomes an effective rescaling of the detection threshold, so there is no loss of generality in supposing $T = 0$.

The prediction of a nonzero dark count rate at zero temperature is, strictly speaking, at variance with quantum mechanical predictions.  Even under ideal conditions, our model predicts a nonzero probability of a vacuum detection event; quantum mechanically this probability should be exactly zero.  However, even at extremely low temperatures, nonzero dark count rates are experimentally observed \cite{Shibata2015}.

For an explicit, albeit notional, example of a physical detection mechanism, one may consider a classical charged particle in a bifurcating harmonic potential.  Such a potential has the quadratic form $\frac{1}{2}m\omega^2 x^2$ for mass $m$ and displacement $x$ for $|x| \le \ell$.  For $|x| > \ell$, the potential is strongly repulsive and the particle quickly accelerates away, thereby creating an observable event.  Since the trapped particle behaves as a high-Q linear filter, its behavior will closely match that of the resonant vacuum mode.  If the polarization is linear and aligned with the displacement of the potential, the particle's motion will bifurcate and run away if the modal amplitude is sufficiently high.

Despite some similarities, the adoption of a threshold detection scheme for modeling photon detection should not be construed as a semi-classical treatment, as we are still completely within the confines of classical physics.  Although we have adopted a very simple model of single-photon detection, these general qualitative observations are expected to hold in a more detailed physical model.  In what follows, we shall make no further reference to the particular physical mechanism used for detection and will instead focus on the more abstract notion of threshold detection in the single-mode limit.


\subsection{Emergence of the Born Rule}

For coherent states, a detection event in the single-mode limit may be written
\begin{equation}
\left| \alpha + \frac{z}{\sqrt{2}} \right|^2 > \gamma^2 \; .
\end{equation}
With detection events so defined, we may identify the complex amplitude $a$, given by
\begin{equation}
a = \alpha + \frac{z}{\sqrt{2}} \; ,
\end{equation}
and note that $4|a|^2$ follows a non-central $\chi^2$ distribution with two degrees of freedom and a noncentrality parameter of $4|\alpha|^2$.  Thus, the cumulative distribution function (cdf) of $|a|^2$ is given by the expression \cite{Johnson1994}
\begin{equation}
\Pr\Bigl[ |a|^2 \le \gamma^2 \Bigr] = 1 - Q_1\left( 2|\alpha|, 2\gamma \right) \; ,
\end{equation}
where $Q_1(\cdot,\cdot)$ is the Marcum Q-function, defined by
\begin{equation}
Q_1(\mu,\nu) = \int_{\nu}^{\infty} x \, e^{-(x^2+\mu^2)/2} \, I_0(\mu x) \, dx \; ,
\end{equation}
and $I_0(\cdot)$ is the zeroth-order modified Bessel function of the first kind.

The probability distribution for $|a|^2$ can be related to the familiar Poisson distribution of photon number in coherent states as follows.  For integers $k \ge 1$, the moments $\mathsf{E}[|a|^{2k}]$ all exist and, therefore, uniquely determine the probability distribution of $|a|^2$.  We observe that the number operator $\hat{n} = \hat{a}^\dagger \hat{a}$ provides a quantum mechanical analog of $|a|^2$ in the sense that $\bra{\alpha} \mathcal{S}(\hat{n}^k) \ket{\alpha} = \mathsf{E}[|a|^{2k}]$, where $\mathcal{S}(\hat{n}^k)$ is the symmetrized form (or Weyl ordering) of $\hat{n}^k$, by the optical equivalence theorem \cite{Cahill1969}.  Furthermore, $\mathcal{S}(\hat{n}^k)$ can be written as a degree-$k$ polynomial $\mathcal{P}_k(\hat{n})$ in $\hat{n}$; for example, $\mathcal{P}_1(\hat{n}) = \hat{n} + \frac{1}{2}$, $\mathcal{P}_2(\hat{n}) = \hat{n}^2 + \hat{n} + \frac{1}{2}$, etc.\ \cite{Fujii2004}.  In terms of the photon number basis, then, we may write the moments as
\begin{equation}
\mathsf{E}[|a|^{2k}] = \bra{\alpha} \mathcal{S}(\hat{n}^k) \ket{\alpha} = \sum_{n=0}^{\infty} \mathcal{P}_k(n) \, |\braket{n|\alpha}|^2 \; ,
\end{equation}
where $|\braket{n|\alpha}|^2 = e^{-|\alpha|^2} |\alpha|^{2n}/n!$ is the probability associated with photon number $n$.  The discrete Poisson distribution for photon number could therefore be interpreted as a calculational device, a mathematical artifice, so to speak, for determining the distribution of the continuous random variable $|a|^2$.  The begs the question of whether threshold exceedances may be interpreted as photon detection events.

The probability of such a detection event is given by
\begin{equation}
\begin{split}
\Pr\left[ |a|^2 > \gamma^2 \right] &= Q_1\left( 2|\alpha|, 2\gamma \right) = \int_{2\gamma}^{\infty} x e^{-(x^2+4|\alpha|^2)/2} \, \frac{1}{\pi} \int_{0}^{\pi} e^{2|\alpha|x\cos\theta} d\theta \, dx \; .
\label{eqn:Marcum}
\end{split}
\end{equation}
We now note that, to fourth order in $|\alpha|$,
\begin{equation}
\begin{split}
\Pr\left[ |a|^2 > \gamma^2 \right] &= \int_{2\gamma}^{\infty} x e^{-x^2/2} \left[ 1+(x^2-2) |\alpha|^2 + \tfrac{1}{4}(x^4-8x^2+8)|\alpha|^4 \right] dx + \mathcal{O}(|\alpha|^6) \\
&= e^{-2\gamma^2} \left( 1 + 4\gamma^2 |\alpha|^2 + 4\gamma^2(\gamma^2-1) |\alpha|^4 \right) + \mathcal{O}(|\alpha|^6) \; .
\label{eqn:approxBorn}
\end{split}
\end{equation}
The presence of $|\alpha|^2$ is the lowest-order approximation is the first indication of the emergence of the Born rule, although the correspondence is subtle and requires some discussion.

According to quantum mechanics, the probability of observing $n$ photons given a coherent state $\ket{\alpha}$ is $p_n = |\braket{n|\alpha}|^2 = e^{-|\alpha|^2} |\alpha|^{2n}/n!$.  Hence, the probability of observing no photons at all is $p_0 = e^{-|\alpha|^2}$, while the probability of observing at least one photon is $1-p_0 = 1-e^{-|\alpha|^2} \approx |\alpha|^2$ for  $|\alpha| \ll 1$.  According to equation (\ref{eqn:Marcum}), for $\alpha = 0$ we have $\Pr[|a|^2 > \gamma^2] = e^{-2\gamma^2}$, which we interpret as the dark count probability of the vacuum state.  For $\alpha \neq 0$, equation (\ref{eqn:approxBorn}) will be a good approximation for $|\alpha|^2 \ll 1/(4\gamma^2)$.  Furthermore, for $\gamma^2 \gg \frac{1}{2}$ we will have low dark counts.  So, for $|\alpha|^2 \ll 1/(4\gamma^2) \ll \frac{1}{2}$ (i.e., $|\alpha|$ small and $\gamma$ large) we expect to be in the near-single-photon regime.  However, taking both $|\alpha|$ to be small and $\gamma$ to be large does not necessarily provide the best agreement with the Born rule, as we shall see.

\begin{figure}[ht]
\centerline{\scalebox{0.75}{\includegraphics{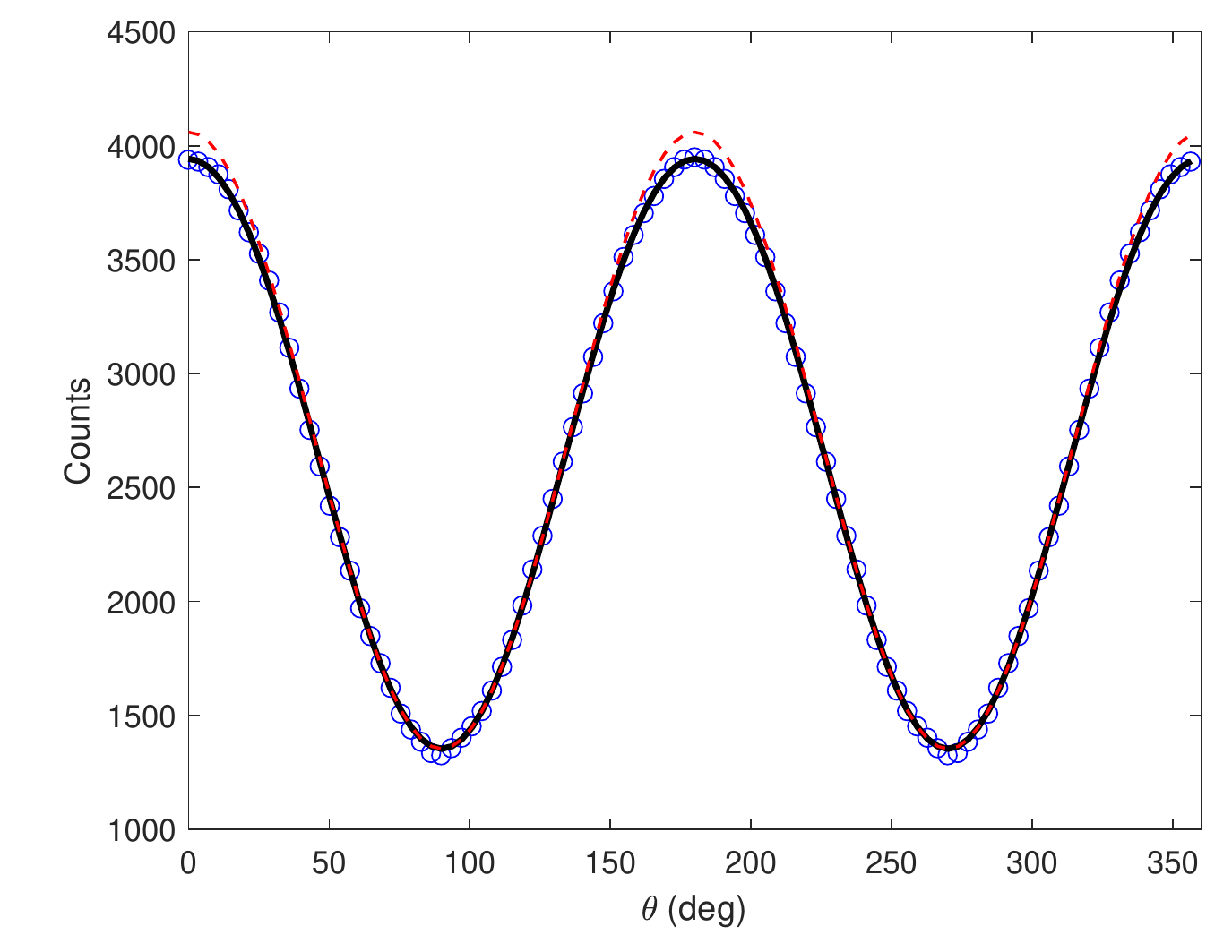}}}
\caption{Plot of simulated counts (blue circles) for $N = 10^4$ trials versus the linear polarization angle $\theta$ for $\alpha = 0.707 \cos\theta$ and $\gamma = 1$.  The detection probability (thick solid black line) given by equation (\ref{eqn:Marcum}) and its approximation (thin dashed red line) given by equation (\ref{eqn:approxBorn}), both scaled by $N$, are shown as well.}
\label{fig:counts}
\end{figure}

Figure \ref{fig:counts} shows an example using $\alpha = 0.707 \cos\theta$, $\gamma =  1$, and $N = 10^4$ random realizations.  Examining $N \Pr[|a|^2 > \gamma^2]$ as a function of $\theta$, we observe a near-perfect sinusoidal pattern with a period of $\pi$ that has a minimum of $N e^{-2\gamma^2} \approx 1353$ and a maximum of $N Q_1(1.414, 2) \approx 3942$.  Subtracting the dark counts and renormalizing by the resulting maximum value, as one normally does in practice, gives a good approximation to the $\cos^2\theta$ probability law one would expect for an application of the Born rule to single-photon detection.  Furthermore, reducing the magnitude of $\alpha$, and of course ignoring the many non-detection events, gives arbitrarily good agreement.  (If $\alpha$ is identically zero we will have a constant dark count rate which, when subtracted out, gives the quantum mechanical prediction of zero.)

\begin{figure}[ht]
\centerline{\scalebox{0.75}{\includegraphics{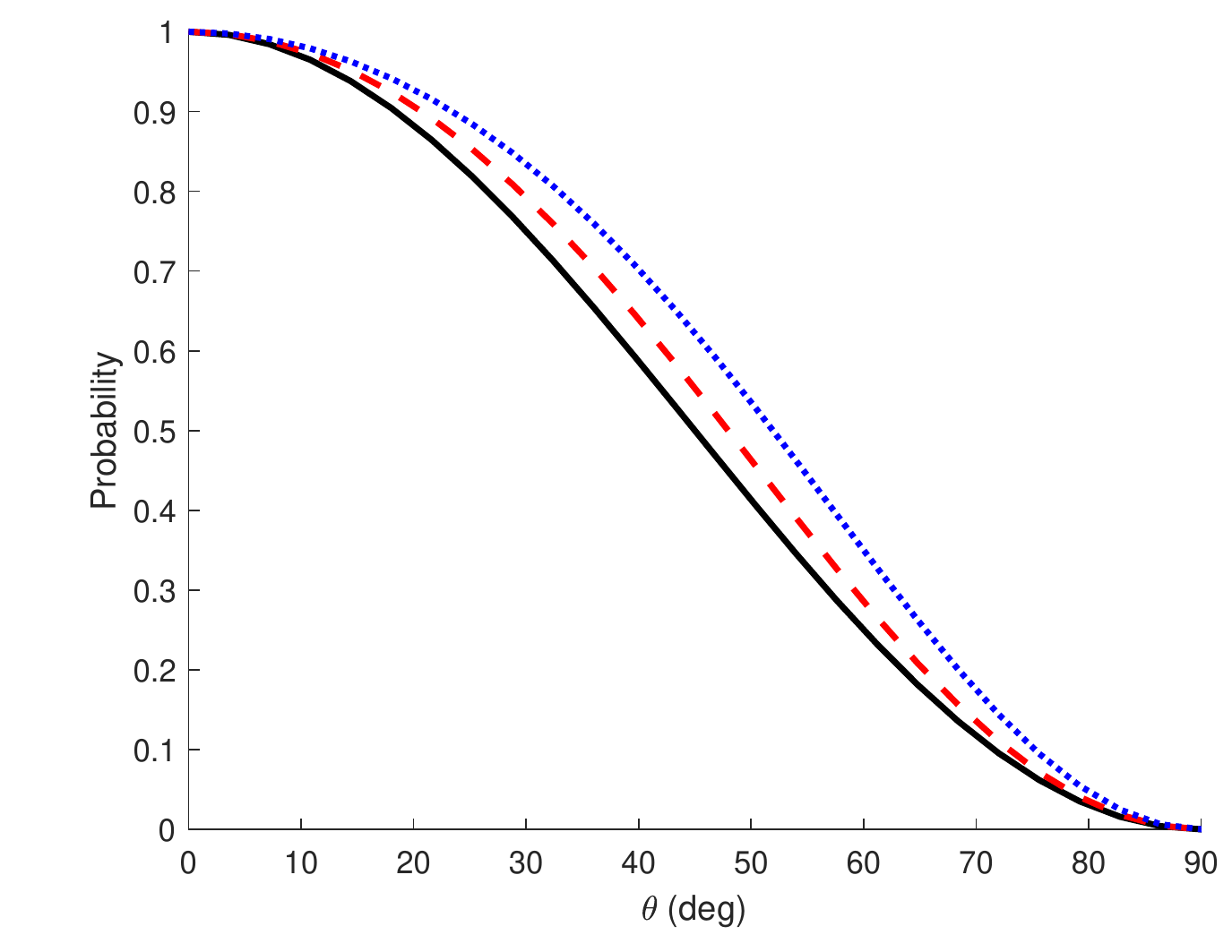}}}
\caption{Plot of predicted probabilities for a coherent state with $\alpha = \cos\theta$ and $\gamma = 1$.  The black solid line is the baseline $\cos^2\theta$ prediction.  The dashed red line is our prediction based on a normalized version of equation (\ref{eqn:Marcum}).  The dotted blue line is the quantum mechanical prediction for detecting one or more photons.}
\label{fig:deviation}
\end{figure}

Now, for a general coherent state $\ket{\alpha}$, quantum mechanics does not actually predict a probability of $\cos^2\theta$, as our detector only indicates the presence of one \emph{or more} photons.  The actual predicted probability is $1-e^{-|\alpha|^2}$, which is only approximately sinusoidal.  Comparing this to equation (\ref{eqn:Marcum}), suitably normalized, we observe a subtle difference.  For $\alpha = \cos\theta$ and $\gamma = 1$, our model predicts a slightly lower probability than the quantum prediction of $1-e^{-|\alpha|^2}$.  (See figure \ref{fig:deviation}.)  For $\gamma = 0.5$ it is slightly higher.  Treating $\gamma$ as an adjustable parameter, then, allows for an arbitrarily good fit.

The Poissonian nature of detection events is also an important characteristic of coherent light.  Strictly speaking, the single-mode approximation we have made entails an infinite coherence time, so the temporal distribution of events is not well defined within the present model.  Under realistic conditions, however, there will be some non-zero bandwidth associated with any given mode and, hence, an associated coherence time $\tau$ inversely proportional to this bandwidth.  For laser light, this is typically on the order of nanoseconds to microseconds.  We may therefore suppose that over a time $T \gg \tau$, there are about $N = T/\tau$ independent detection opportunities.  Under our model, the probability of a detection for each such opportunity, after subtracting for dark counts and assuming $\gamma |\alpha| \ll 1$, is about $p = 4\gamma^2 e^{-2\gamma^2} |\alpha|^2$.  Now, the number of detections out of these $N$ trials is binomially distributed with a mean of $Np$ and a variance of $Np(1-p)$.  If $N$ is large and $p$ is small (e.g., $|\alpha| \ll 1$ for fixed $\gamma$), this distribution becomes approximately Poissonian.  The average number of counts, $Np$, may be interpreted as the product of an incident photon rate $|\alpha|^2/\tau$, a detection efficiency $4\gamma^2 e^{-2\gamma^2}$, and an observation time $T$.  Deviations from Poissonian behavior are to be expected when $|\alpha|$ is large, the detection efficiency is high, or the observation time is short.

A further comparison to experimental observations can be made by considering detection efficiency.  For Poisson-distributed photon statistics, experimentalists often use a parametric model of the form
\begin{equation}
p = 1 - (1-\delta) e^{-\eta|\alpha|^2} \; ,
\label{eqn:exBorn}
\end{equation}
where $p$ is the probability of a detection event, $\delta$ is the dark count probability, and $\eta \in [0,1]$ is the detection efficiency \cite{Campbell1984,Oh2010}.  Our model conforms with this general expression in the small $|\alpha|$ limit if we take $\delta = e^{-2\gamma^2}$ and 
\begin{equation}
\eta = \frac{4\gamma^2 \, e^{-2\gamma^2}}{1-e^{-2\gamma^2}} \; .
\end{equation}
Note that, in this interpretation, the effective detection efficiency increases as the threshold $\gamma$ is decreased, attaining near unit efficiency for $\gamma \approx 0.8$; however, for values much lower than this the efficiency is over unity and this interpretation is no longer valid.

\begin{figure}[ht]
\centerline{\scalebox{0.75}{\includegraphics{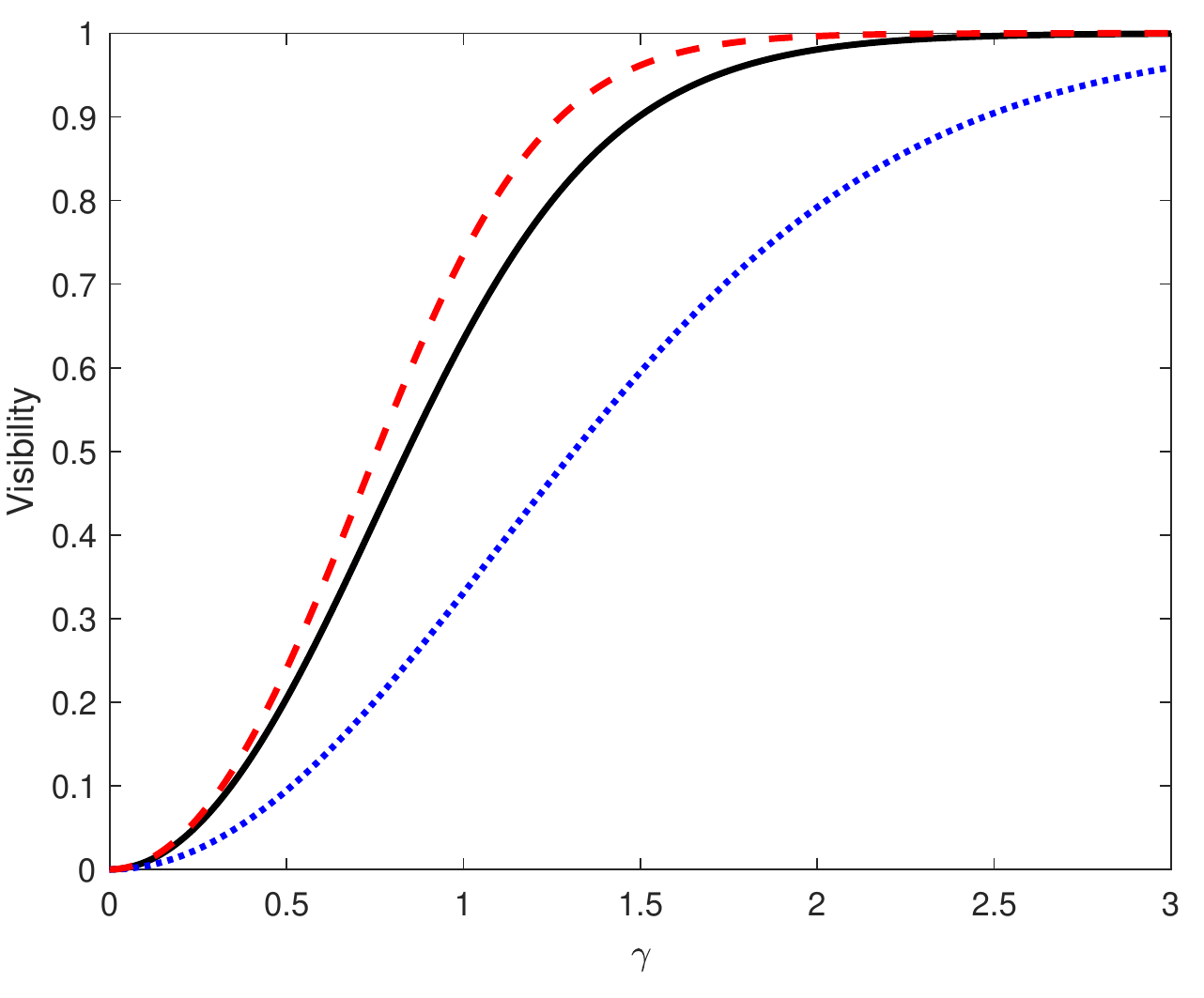}}}
\caption{Plot of the interferometric visibility as a function of the detection threshold $\gamma$.  The lower blue dotted curve is for $\alpha = 0.5$, the middle black solid curve is for $\alpha = 1$, and the upper red dashed curve is for $\alpha = 1.5$.}
\label{fig:visibility}
\end{figure}

Finally, another important quantity in experimental quantum optics is the interferometric visibility, which measures the degree of coherence in the prepared state.  This may be defined as the ratio of the difference in maximum and minimum probabilities to their sum, which in our case is
\begin{equation}
\mathcal{V} = \frac{Q_1(2|\alpha|,2\gamma) - e^{-2\gamma^2}}{Q_1(2|\alpha|,2\gamma) + e^{-2\gamma^2}} \; .
\end{equation}
Taking $\gamma$ to be large, with $\alpha$ fixed, therefore gives a fringe visibility arbitrarily close to unity.  For larger values of $|\alpha|$, corresponding more closely to the classical regime, the convergence to unity occurs more rapidly.  We illustrate this in figure \ref{fig:visibility}, plotting visibility as a function of the threshold for different values of $\alpha$.  It is important to note that the visibility described here is in terms of the probability of detection, $\Pr[|a|^2 > \gamma^2]$, not the intensity, $|a|^2$, which is random, nor the expected intensity, $\mathsf{E}[\,|a|^2]$, which would give a visibility of one half.  This point is important for a proper comparison with quantum mechanics, which predicts visibilities as high as one for actual measured counts, not classical intensities.

In summary, we expect to find good agreement with quantum mechanical predictions when $|\alpha|$ is small and $\gamma$ is large, with optimal but imperfect agreement for suitable finite, nonzero choices of both.  Parameter regimes in which agreement with the Born rule is best, such as low values of $|\alpha|$, may produce lower visibility.  In general, no combination of $\alpha$ and $\gamma$ can give both arbitrarily good agreement with the Born rule and a visibility arbitrarily close to unity.  The detailed tradeoffs between these and other competing metrics are discussed further in section \ref{sec:GLT}.


\subsection{Dual-mode Detection}

Previously, we considered measurements along a single polarization mode and found that the associated probabilities follow the Born rule, albeit with a threshold-dependent rescaling and fixed offset in accordance with equation (\ref{eqn:exBorn}).  Such measurements cannot distinguish between a missed detection and an event that would have resulted in a detection in an orthogonal polarization.  Dual-modal detection provides an alternative method for comparing against the Born rule that overcomes this deficiency.

Let $\unit{e}_H$ and $\unit{e}_V$ denote the horizontal and vertical polarization unit vectors for given wave vector.  A linearly polarized coherent state for this wave vector may be described by the complex amplitude vector
\begin{equation}
\vec{a} = a_H \unit{e}_H + a_V \unit{e}_V = \left( \alpha\cos\theta + \frac{z_H}{\sqrt{2}} \right) \unit{e}_H + \left( \alpha\sin\theta + \frac{z_V}{\sqrt{2}} \right) \unit{e}_V \; ,
\end{equation}
where $z_H$ and $z_V$ are independent standard complex Gaussian random variables.  Note that $\theta=0$ corresponds to a vacuum state in the orthogonal polarization mode, which is always assumed to be present.

As a consequence of the independence of $z_H$ and $z_V$, the random variables $|a_H|^2$ and $|a_V|^2$ are also independent, and their joint cdf is given by the product of their marginal distributions.  Let us suppose a dual-mode detector that will register separate events if either $|a_H| > \gamma$ or $|a_V| > \gamma$ are true.  This would be the case if the detector were, say, a pair of bifurcating harmonic oscillators oriented in the horizontal and vertical polarization directions.  Equivalently, we may consider a polarizing beam splitter that separates the components to two single-mode detectors.  The probability of no detection occurring is then
\begin{equation}
\begin{split}
P_{0} &= \Pr\!\left[ |a_H|^2 \le \gamma^2, |a_V|^2 \le \gamma^2 \right] \\
&= \Bigl[ 1 - Q_1(2|\alpha\cos\theta|, 2\gamma) \Bigr] \Bigl[ 1 - Q_1(2|\alpha\sin\theta|, 2\gamma) \Bigr]
\end{split}
\end{equation}
Likewise, the probabilities for the three possible detection events are
%
\begin{eqnarray}
P_{H} &= Q_1(2|\alpha\cos\theta|, 2\gamma) \Bigl[ 1 - Q_1(2|\alpha\sin\theta|, 2\gamma) \Bigr] \\
P_{V} &= \Bigl[ 1 - Q_1(2|\alpha\cos\theta|, 2\gamma) \Bigr] Q_1(2|\alpha\sin\theta|, 2\gamma) \\
P_{HV} &= Q_1(2|\alpha\cos\theta|, 2\gamma) \, Q_1(2|\alpha\sin\theta|, 2\gamma) \; ,
\end{eqnarray}
%
where $P_{H}$ is the probability of a single detection of $H$, $P_{V}$ is the probability of a single detection of $V$, and $P_{HV}$ is the probability of both.

In actual experiments with coherent light it is common to reject events in which there are two detections and, of course, ignore those with none.  Out of a notional, unknown number $N$ of independent trials, one measures $S_H = N P_{H}$ single counts for $H$, $S_V = N P_{V}$ single counts for $V$, and $N P_{HV}$ ``accidental'' coincidence counts.  If we post-select on single detection events, the conditional probability, $p_H$, of detecting $H$ is
\begin{equation}
p_H = \frac{S_H}{S_H + S_V} = \frac{P_{H}}{P_{H}+P_{V}} \; .
\label{eqn:probH}
\end{equation}

\begin{figure}[ht]
\centerline{\scalebox{0.75}{\includegraphics{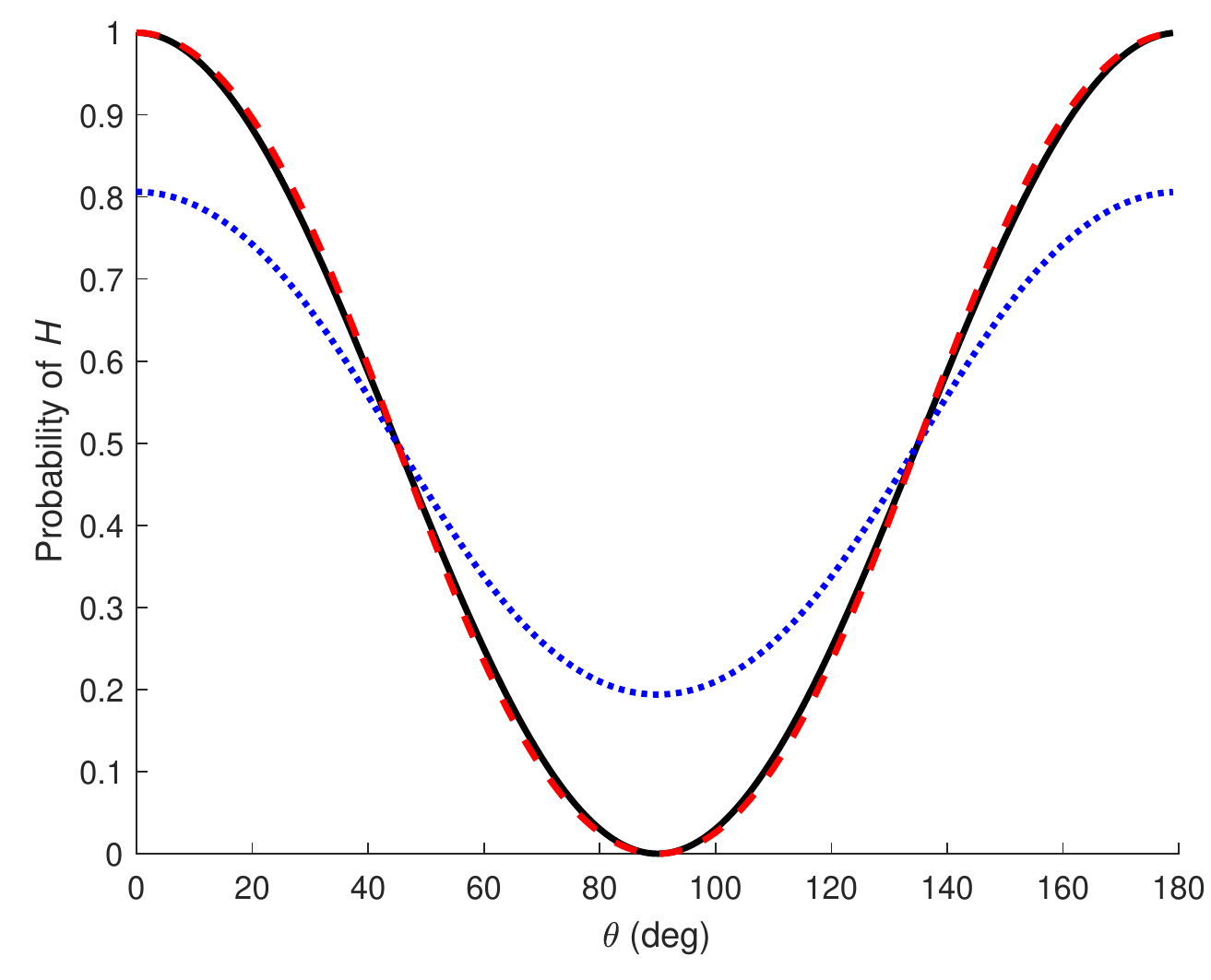}}}
\caption{Plot of the conditional probability $p_H$ (blue dotted line) and renormalized conditional probability $\hat{p}_H$ (red dashed line) against the $\cos^2\theta$ Born rule prediction (black solid line), versus the polarization angle $\theta$ for $|\alpha|^2 = 0.5$ and $\gamma = 1$.}
\label{fig:BornAgain}
\end{figure}

We may now compare $p_H$ with the Born rule prediction of $\cos^2\theta$.  An example is plotted in figure \ref{fig:BornAgain} for $|\alpha|^2 = 0.5$ and $\gamma = 1$.  The agreement is perfect when $\theta = 45^\circ, 135^\circ$ (diagonal and anti-diagonal polarization, respectively), resulting in balanced probabilities and a conditional probability of $\frac{1}{2}$.  For other values of $\theta$, we find $\frac{1}{2}(1-\mathcal{V}) \le p_H \le \frac{1}{2}(1+\mathcal{V})$, where
\begin{equation}
\mathcal{V} = \frac{Q_1(2|\alpha|,2\gamma) (1-e^{-2\gamma^2}) - [1-Q_1(2|\alpha|,2\gamma)] e^{-2\gamma^2}}{Q_1(2|\alpha|,2\gamma)(1-e^{-2\gamma^2}) + [1-Q_1(2|\alpha|,2\gamma)] e^{-2\gamma^2}}
\end{equation}
is the visibility.  For our particular case, $\mathcal{V} = 0.61$, so $0.19 \le p_H \le 0.81$.

The maximum discrepancy arises when $\theta$ is $0^\circ$ or $90^\circ$.  For $\theta = 0^\circ$, the polarization of the wave is horizontal, but we are still not guaranteed an $H$ outcome, even conditionally, because the probability of a ``false'' $V$ detection is still nonzero.  Similarly, for $\theta = 90^\circ$, the polarization of the wave is vertical, but an $H$ outcome is still possible due to dark counts.  In any realistic experiment, such events will be unavoidable and are quantified by a visibility below unity.  Such anomalous events are effectively removed by renormalization, resulting in a modified conditional probability of the form
\begin{equation}
\hat{p}_H = \frac{1}{\mathcal{V}} \left( p_H - \frac{1}{2} \right) + \frac{1}{2} \; .
\end{equation}
This renormalized conditional probability gives excellent agreement with the Born rule prediction, as shown in figure \ref{fig:BornAgain}.

\subsection{Particle-like Behavior}

Consider a coherent state prepared in some polarization mode $\unit{e}_0$ and spatial mode $\vec{k}_{R}$ traveling to the right  that is incident upon a 50/50 beam splitter (BS).  The outgoing beams have orthogonal spatial modes of $\vec{k}_{R}$ and $\vec{k}_{D}$ traveling right and down, respectively, each with the same polarization mode.  The initial state may be described by the vector
\begin{equation}
\vec{a} = \begin{bmatrix} \alpha + z_{R}/\sqrt{2} \\ z_{D}/\sqrt{2} \end{bmatrix} = \alpha \begin{bmatrix} 1 \\ 0 \end{bmatrix} + \frac{1}{\sqrt{2}} \begin{bmatrix} z_{R} \\ z_{D} \end{bmatrix} \; ,
\end{equation}
where $z_{R}$ and $z_{D}$ are independent standard complex Gaussian random variables corresponding to the ZPF components of the two spatial modes.  For simplicity, we ignore the orthogonal polarization modes.

The beam splitter acts as a Hadamard gate $\mtx{H}$, transforming $\vec{a}$ into
\begin{equation}
\vec{a}' = \mtx{H} \vec{a} = \frac{\alpha}{\sqrt{2}} \begin{bmatrix} 1 \\ 1 \end{bmatrix} + \frac{1}{2} \begin{bmatrix} z_{R} + z_{D} \\ z_{R} - z_{D} \end{bmatrix} \; .
\end{equation}
Note that $z_{R}' = (z_{R} + z_{D})/\sqrt{2}$ and $z_{D}' = (z_{R} - z_{D})/\sqrt{2}$ are again independent standard complex Gaussian random variables, so the noise term for $\vec{a}'$ has the same form as that for $\vec{a}$.

If we place single-mode detectors at each output port of the beam splitter, there will be four possible outcomes with four corresponding probabilities: no detections ($P_{0}$), a single detection for mode $\vec{k}_{R}$ ($P_{R}$), a single detection for mode $\vec{k}_{D}$ ($P_{D}$), and coincident detections on both modes ($P_{RD}$).  These probabilities are as follows:
\begin{align}
P_{0} &= \Bigl[ 1 - Q_1(\sqrt{2}|\alpha|, 2\gamma) \Bigr]^2 \\
P_{R} &= P_{D} = \Bigl[ 1 - Q_1(\sqrt{2}|\alpha|, 2\gamma) \Bigr] Q_1(\sqrt{2}|\alpha|, 2\gamma) \\
P_{RD} &= Q_1(\sqrt{2}|\alpha|, 2\gamma)^2 \; .
\end{align}

Note that $P_{RD} \ge P_{R} P_{D}$, since each detection event is independent of the other.  A similar result is found in the semiclassical treatment of photon detection \cite{Loudon1980}.  In the single-photon regime ($|\alpha| \ll 1$) one would expect particle-like behavior, so coincident detections should be quite rare.  Quantum mechanically, the probability of a coincident detection for a true, single-photon state would be exactly zero.

Experimentally, one counts the number of single-detection events, $S_{R} = N P_{R}$ and $S_{D}= N P_{D}$, for transmitted and reflected light, respectively, as well as the number of coincidences, $C = N P_{RD}$, where $N$ is the nominal number of trials.  The difficulty with such experiments is that $N$ is often unknown or perhaps unknowable.  If $N$ is known, the ratio $R = C N/(S_{R} S_{D})$, more commonly associated with the degree of second-order temporal coherence $g^{(2)}(0)$, would be expected to have a value no less than one, since
\begin{equation}
R = \frac{C N}{S_{R} S_{D}} = \frac{P_{RD}}{P_{R} P_{D}} \ge 1 \; .
\end{equation}
If $C = 0$, as predicted by quantum mechanics, and $S_{R}, S_{D} > 0$, then $R = 0$, thereby violating the inequality.  Early experiments of this sort were performed by Grangier \textit{et al.} using both a light-emitting diode (LED) \cite{Grangier2019} and a heralded photon source \cite{Grangier1986}.  The LED light source was turned on briefly using a controlled electronic trigger, allowing $N$ to be know precisely.  Since the LED light was strongly attenuated, a value of $R$ near unity, and consistent with the inequality $R \ge 1$, was measured, as one might expect.

In the case of the heralded photon source, $N$ was taken to be the number of trigger events, $N_t$, each of which was taken to indicate the presence of a single, heralded photon.  Under this assumption, the experimenters obtained a value of $R_t = C N_t/(S_{R} S_{D})$ significantly less than one.  A value less than one is generally considered to be evidence of photon antibunching.  The true value of $N$, however, could not be known and may well have been much larger than $N_t$, in which case a value below unity would not be surprising.  A similar experiment, also using heralded events, was performed recently by Thorn \textit{et al.}\ using a modern parametric downconversion source and avalanche photodiodes, with similar results \cite{Thorn2004}.

\begin{figure}[ht]
\centerline{\scalebox{0.75}{\includegraphics{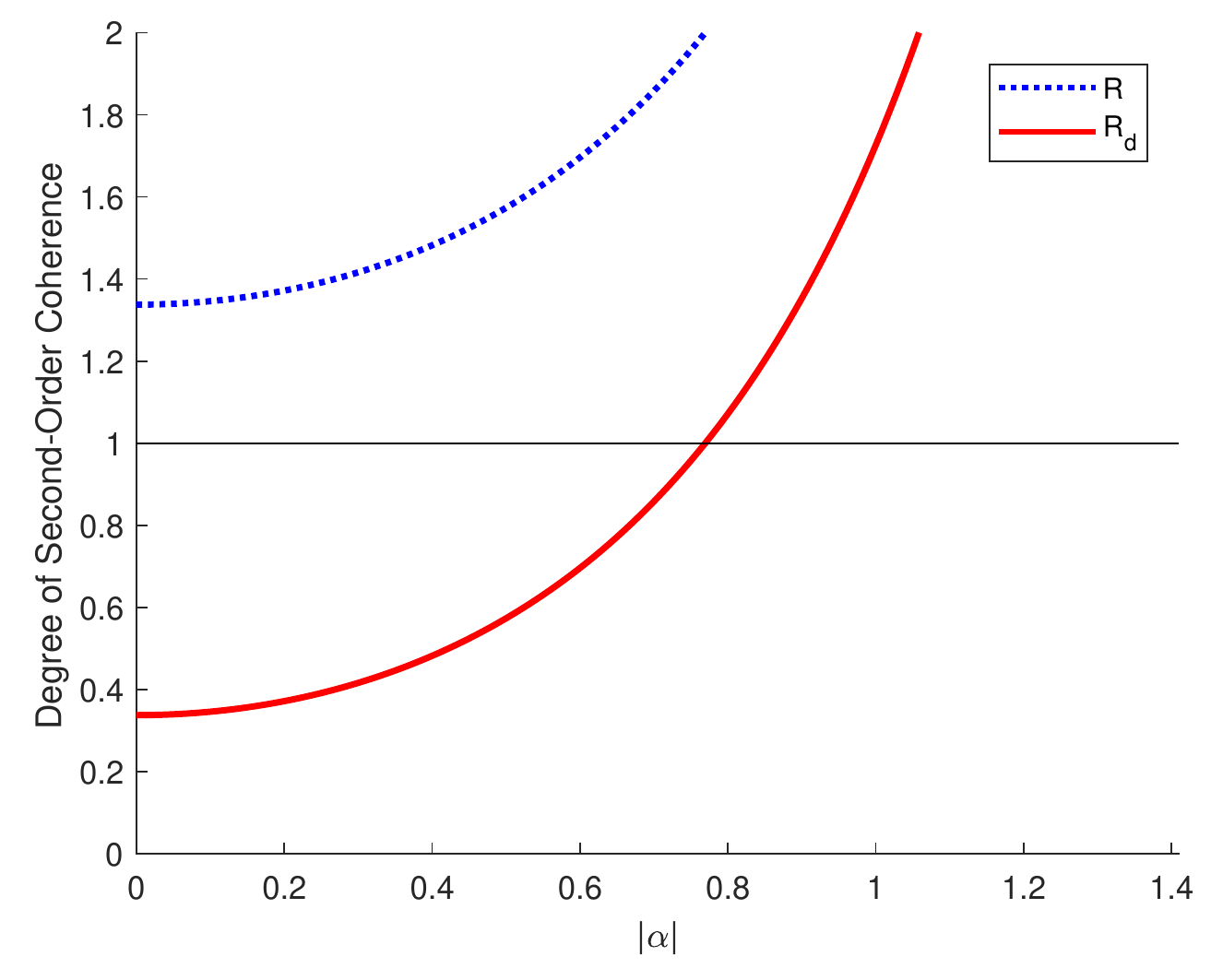}}}
\caption{Plot of $R$ (dotted blue line) and $R_d$ (solid red line) as a function of $|\alpha|$ for $\gamma = 1$.  Values below one indicate particle-like behavior.}
\label{fig:antibunching}
\end{figure}

For our model, the single-photon regime provides a good approximation to a true, single-photon state, so long as we ignore non-detection events.  Taking $N$ to be $N_d = S_{R} + S_{D} + C$, the total number of detection events, we obtain a result similar to heralding.  From this we may compute the ratio
\begin{equation}
R_d = \frac{C N_d}{S_{R} S_{D}} = \frac{P_{RD} (1-P_{0})}{P_{R} P_{D}} \; .
\end{equation}
This may equivalently be seen as replacing the absolute probabilities $P_{R}, P_{D}, P_{RD}$  in the expression for $R$ with the \emph{conditional} probabilities $p_{R} = P_{R}/(1-P_{0}), p_{D} = P_{D}/(1-P_{0}), p_{RD} = P_{RD}/(1-P_{0})$.  Such conditioning is similar to the experimental procedure of using heralding to define the number of trials.  It is now easy to show that $R_d$ can be less than unity when either $|\alpha| \ll 1$ or $\gamma \gg 1$.  As an example, figure \ref{fig:antibunching} shows the values of $R$ and $R_d$ as a function of $|\alpha|$ for $\gamma = 1$.  For this example, $R_d$ achieves a minimum value of about 0.34, whereas quantum mechanics predicts $R = 0$ for an ideal single-photon state.  By increasing the value of $\gamma$, however, this minimum can be made arbitrarily small.  For example, the observed value of $R_d = 0.018$ in reference \cite{Thorn2004} could be achieved with $\alpha = 0.3$ and $\gamma = 1.6$.  In this way, a purely classical model of light, when analyzed in a similar way, can exhibit the same anomalous quantum behavior.

\section{General Linear Transformations}
\label{sec:GLT}


\subsection{Single-Photon, Four-Mode Entanglement}

In quantum mechanics, a single photon can be entangled across multiple modes.  Similar behavior can be modeled classically.  Consider a coherent state prepared with polarization $\unit{e}_{H}$ traveling to the right and incident upon a 50/50 beam splitter (BS).  The initial, four-mode state may be written
\begin{equation}
\vec{a} = \begin{bmatrix} \left( \alpha + \frac{z_1}{\sqrt{2}} \right) \unit{e}_{H} + \frac{z_2}{\sqrt{2}} \unit{e}_{V} \\ \frac{z_3}{\sqrt{2}} \unit{e}_{H} + \frac{z_4}{\sqrt{2}} \unit{e}_{V} \end{bmatrix}
= \alpha \begin{bmatrix} 1 \\ 0 \\ 0 \\ 0 \end{bmatrix} + \frac{1}{\sqrt{2}} \begin{bmatrix} z_{RH} \\ z_{RV} \\ z_{DH} \\ z_{DV} \end{bmatrix}
\end{equation}
where $z_{RH}, z_{RV}, z_{DH}, z_{DV}$ are independent and identically distributed (iid) standard complex Gaussian random variables arising from the zero-point field and corresponding to the two spatial modes ($\vec{k}_{R}$ and $\vec{k}_{D}$) and polarization modes ($\unit{e}_H$ and $\unit{e}_V$).

After the beam splitter, the state becomes
\begin{equation}
\vec{a}' = ( \mtx{H} \otimes \mtx{I} ) \vec{a} = \frac{\alpha}{\sqrt{2}} \begin{bmatrix} 1 \\ 0 \\ 1 \\ 0 \end{bmatrix} + \frac{1}{2} \begin{bmatrix} z_{RH} + z_{DH} \\ z_{RV} + z_{DV} \\ z_{RH} - z_{DH} \\ z_{RV} - z_{DV} \end{bmatrix} \; ,
\end{equation}
where $\otimes$ is the Kronecker product and $\mtx{I}$ is the $2 \times 2$ identity.  Finally, we may apply an $\mtx{X}$ gate (i.e., a half-wave plate rotated $45^\circ$) on the downward mode to change the polarization.  This has the effect of performing a controlled NOT gate $\mtx{C}$, with the spatial mode (i.e., the optical path) as the control and the polarization as the target.  The resulting state is now
\begin{equation}
\vec{a}'' = \mtx{C} ( \mtx{H} \otimes \mtx{I} ) \vec{a} = \frac{\alpha}{\sqrt{2}} \begin{bmatrix} 1 \\ 0 \\ 0 \\ 1 \end{bmatrix} + \frac{1}{2} \begin{bmatrix} z_{RH} + z_{DH} \\ z_{RV} + z_{DV} \\ z_{RV} - z_{DV}  \\ z_{RH} - z_{DH} \end{bmatrix}
\end{equation}
Note that the second term is again a vector of iid standard complex Gaussian random variables.  We may therefore rewrite $\vec{a}''$ as 
\begin{equation}
\vec{a}'' = \frac{\alpha}{\sqrt{2}} \begin{bmatrix} 1 \\ 0 \\ 0 \\ 1 \end{bmatrix} + \frac{1}{\sqrt{2}} \begin{bmatrix} z_{RH}' \\ z_{RV}' \\ z_{DH}' \\ z_{DV}' \end{bmatrix} = \alpha \, \vec{\psi} + \frac{\vec{z}'}{\sqrt{2}} \; ,
\end{equation}
where $\vec{z}' = [z_{RH}', \ldots, z_{DV}']^\mathsf{T}$ and $\vec{\psi}$ is a column vector of unit amplitude.  The vector $\vec{\psi}$ has the mathematical form of an entangled Bell state
\begin{equation}
\vec{\psi} = \frac{\ket{R,H} + \ket{D,V}}{\sqrt{2}} \; ,
\end{equation}
where $\ket{R,H} = \ket{R} \otimes \ket{H} = [1,0]^\mathsf{T} \otimes [1,0]^\mathsf{T}$ and $\ket{D,V} = \ket{D} \otimes \ket{V} = [0,1]^\mathsf{T} \otimes [0,1]^\mathsf{T}$.

To perform a measurement of all four modes, each spatial mode is put into a dual-mode detector and threshold detection is performed.  There are four components and, so, 16 possible outcomes, including multiple detections.  For $|\alpha| \ll 1$, the most likely outcome is no detections at all, with single detections being the next most likely outcome.  At the opposite extreme, for $|\alpha| \gg 1$ the most likely outcome is detection on all four modes.  For small values of $|\alpha|$, the probability of a single detection on either $\ket{R,H}$ or $\ket{D,V}$ (both equally likely) is much more likely than a single detection on $\ket{D,H}$ or $\ket{R,V}$.

\begin{figure}[ht]
\centerline{\scalebox{0.70}{\includegraphics{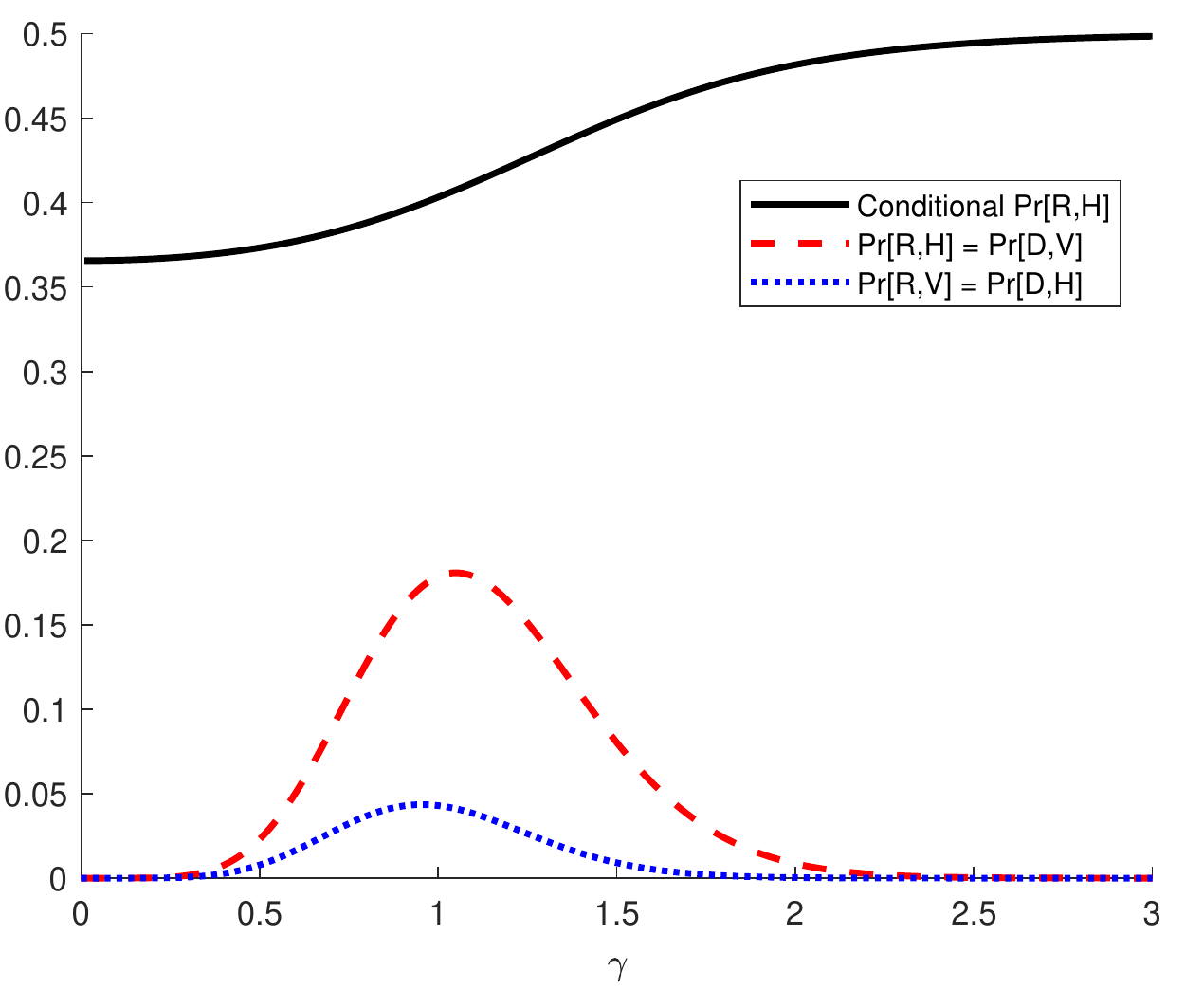}}}
\caption{Plots of $\Pr[R,H] = \Pr[D,V]$ (red dashed line), $\Pr[R,V] = \Pr[D,H]$ (dotted blue line), and the conditional probability of mode $\ket{R,H}$ given any single-mode detection (solid black line) versus $\gamma$ and for $|\alpha| = 1$.}
\label{fig:hyperentanglement}
\end{figure}

Let $\Pr[R,H] = \Pr[D,V]$ and $\Pr[R,V] = \Pr[D,H]$ denote the probabilities for single-detection events on each of the four modes.  These will be given by
\begin{align}
\Pr[R,H] &= \Pr[D,V] = \left( 1-e^{-2\gamma^2} \right)^2 Q_1(\sqrt{2}|\alpha|,2\gamma) \Bigl[ 1-Q_1(\sqrt{2}|\alpha|,2\gamma) \Bigr] \\
\Pr[R, V] &= \Pr[D,H] = e^{-2\gamma^2} \left( 1-e^{-2\gamma^2} \right) \Bigl[ 1-Q_1(\sqrt{2}|\alpha|,2\gamma) \Bigr]^2
\end{align}
These probabilities are illustrated in figure \ref{fig:hyperentanglement} for $\alpha = 1$.  We see that the dominant modes peak in probability at threshold values somewhat greater than $1$ but are relatively much larger than the other two modes.  This comports with the general behavior one would expect of a single-photon state that is hyperentangled in spatial and polarization modes \cite{Kwiat1998}.  If we consider \emph{only} single-mode detection events (i.e., detections on one spatial mode and one polarization mode), then the conditional probability of each dominant mode converges to 0.5, the ideal quantum prediction, when $\gamma$ is large.  Conversely, the conditional probability converges to a nonzero value, which is dependent on $\alpha$, when $\gamma$ is small.  Qualitatively similar behavior is found when $|\alpha|$ is varied while holding $\gamma$ fixed.  Thus, a correspondence with quantum mechanical predictions is achieved, but only in the limit of larger threshold values and only when one post-selects on single-mode detection events.  This asymptotic behavior is a result of the symmetry of the Bell state and would not be expected more generally.  Correlations between the modes are purely a result of post-selection, as the modes themselves are statistically independent.  Of course, this model reproduces only \emph{local} correlations between single-photon modes and not the \emph{nonlocal} correlations one expects from a multi-photon entangled state.

\subsection{Wave/Particle Duality}

In quantum mechanics, photons can exhibit both particle- and wave-like behavior.  This, too, can be modeled classically.  Consider, an initial quantum state $\ket{R,H}$ that undergoes a transformation via a 50/50 beam splitter and a phase shifter in the $\ket{D,H}$ mode.  Using a pair of mirrors, the two paths are recombined in a second beam splitter to form a Mach-Zehnder interferometer.  The two output ports are then measured with detectors.  Quantum mechanically, the final state (before measurement) is
\begin{equation}
\begin{split}
\vec{\psi} &= (\mtx{H} \otimes \mtx{I}) (\mtx{R}_{\phi} \otimes \mtx{I}) (\mtx{H} \otimes \mtx{I}) \ket{R,H} \\
&= \tfrac{1}{2} (1+e^{i\phi}) \ket{R,H} \;+\; \tfrac{1}{2} (1-e^{i\phi}) \ket{D,H} \; ,
\end{split}
\end{equation}
where $\mtx{R}_{\phi}$ is the phase shift gate
\begin{equation}
\mtx{R}_{\phi} = \begin{pmatrix} 1 & 0 \\ 0 & e^{i\phi} \end{pmatrix} \; .
\end{equation}
Accordingly, the probability of finding a photon in the $\ket{R,H}$ mode is $\cos^2(\phi/2)$.

\begin{figure}[ht]
\centerline{\scalebox{0.55}{\includegraphics{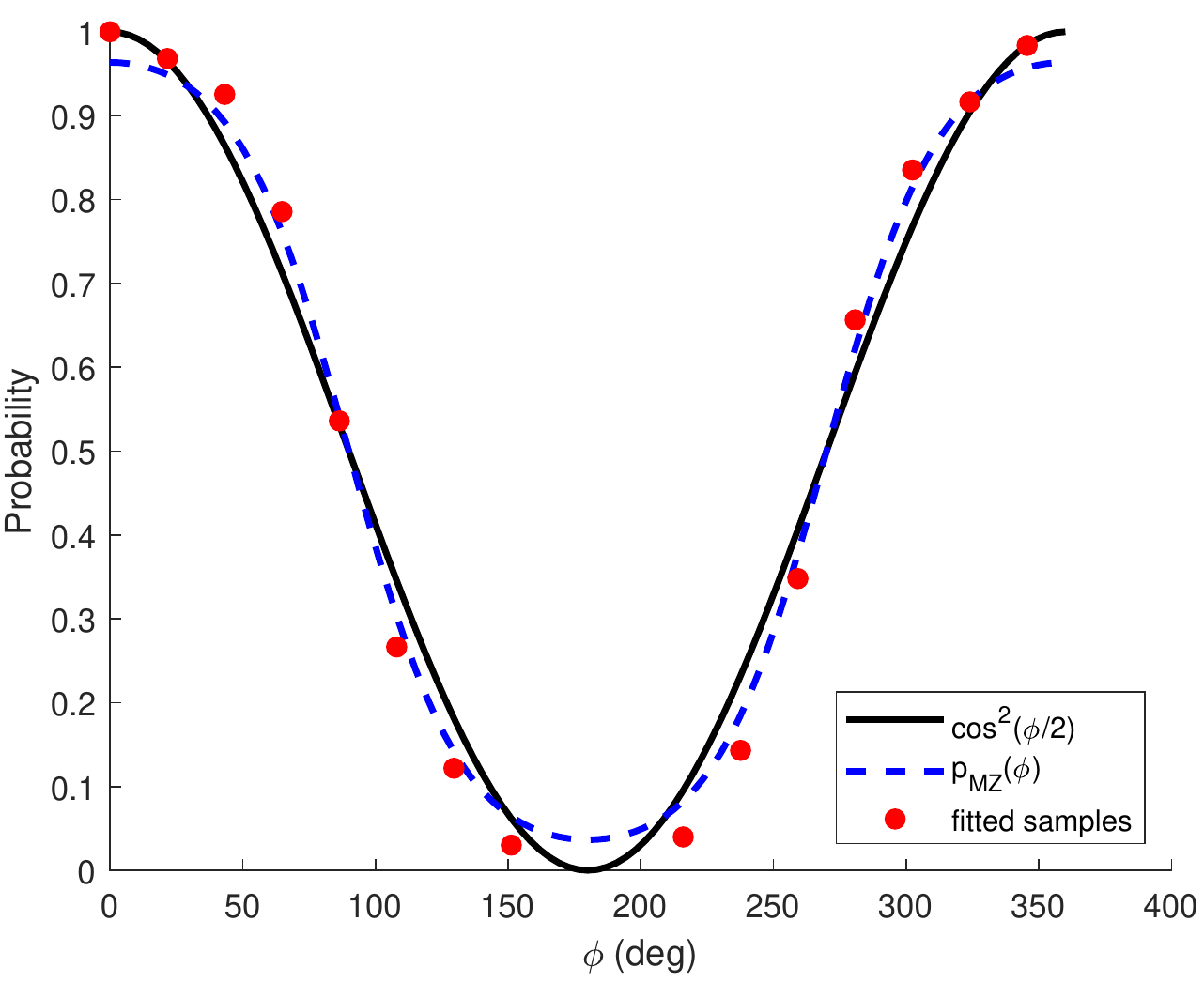}}}
\caption{Plot of the idealized Mach-Zehnder interference pattern given by $\cos^2(\phi/2)$ (black solid line) and the interference pattern predicted by $p_{\rm MZ}(\phi)$ of equation (\ref{eqn:MZ}) (blue dashed line) for $\alpha = 0.95$ and $\gamma = 1.6$.  The red circles are fitted samples analogous to those in reference \cite{Jacques2007}.}
\label{fig:MachZehnderProb}
\end{figure}

We can model the problem classically by starting with an initial coherent state $\vec{a} = \alpha \ket{R,H} + \vec{z}/\sqrt{2}$ and transforming it via the same linear operations into
\begin{equation}
\vec{a}_{\rm MZ} = (\mtx{H} \otimes \mtx{I}) (\mtx{R}_{\phi} \otimes \mtx{I}) (\mtx{H} \otimes \mtx{I}) \vec{a} = \alpha \vec{\psi} + \frac{\vec{z}'}{\sqrt{2}} \; .
\end{equation}
The conditional probability of a detection in mode $\ket{R,H}$, given a single detection in either mode $\ket{R,H}$ or $\ket{D,H}$, is now found to be
\begin{equation}
p_{\rm MZ}(\phi) = \left( 1 + \frac{Q_1(|\alpha(1-e^{i\phi})|,2\gamma)}{Q_1(|\alpha(1+e^{i\phi})|,2\gamma)} \cdot \frac{1-Q_1(|\alpha(1+e^{i\phi})|,2\gamma)}{1-Q_1(|\alpha(1-e^{i\phi})|,2\gamma)} \right)^{-1} \; .
\label{eqn:MZ}
\end{equation}
The resulting interference pattern, as shown in figure \ref{fig:MachZehnderProb}, is similar to what one would expect from classical light if one were observing intensities; however, we are showing probabilities.  The pattern also reflects the nonlocal (i.e., spatially distributed) nature of the interferometer: light travels along both arms and interferes only when recombined.  In this way, our classical model exhibits the wave-like nature of light in terms of discrete detection events.

Figure \ref{fig:MachZehnderProb} also affords an interesting comparison to experimental data.  Reference \cite{Jacques2007} describes an experimental realization of Wheeler's delayed-choice experiment using a nitrogen-vacancy center single-photon source.  The experimenters report an anti-correlation parameter of $R_d = 0.12$ and a visibility of $\mathcal{V} = 94\%$, after subtracting dark counts, when the second beam splitter is in place.  An estimated 2600 photons were used for each measurement sample, and the resulting data points were fit to a cosine wave with a period of $2\pi$.  We have followed a similar analysis procedure in figure \ref{fig:MachZehnderProb}, using a set of fitted samples with a standard deviations of $1/\sqrt{2600}$ and subtracting the expected dark count probability of $e^{-2\gamma^2}$.  Using parameter values of $\gamma = 1.6$ and $\alpha = 0.95$ gives $R_d = 0.12$, $\mathcal{V} = 94\%$, and a root-mean-square error of 0.04, in excellent agreement with the experimental results.  Although the two light sources are quite different, our model is nevertheless capable of reproducing similar observations.

To recover the particle-like nature of the system, we may create a Wheeler delayed-choice experiment by removing the final beam splitter  The resulting state is now
\begin{equation}
\vec{a}_{\rm DC} = (\mtx{R}_{\phi} \otimes \mtx{I}) (\mtx{H} \otimes \mtx{I})  \vec{a} = \frac{\alpha}{\sqrt{2}} \Bigl( \ket{R,H} + e^{i\phi} \ket{D,H} \Bigr) + \frac{\vec{z}''}{\sqrt{2}} \; .
\end{equation}
The conditional probability of a detection in mode $\ket{R,H}$ given a single photon detection in either mode $\ket{R,H}$ or $\ket{D,H}$ is now $\frac{1}{2}$, independent of $\phi$.  In addition, the probability of a double detection in both modes may be made arbitrarily small by decreasing $|\alpha|$ or, equivalently, increasing $\gamma$.  This is the behavior one would expect from a localized particle.  Note that it does not matter when the choice to remove the final beam splitter is made.

A similar result is obtained if we simply provide ``which way'' information by marking one of the two arms with, say, a change in polarization.  If we apply an $\mtx{X}$ gate on the lower arm before the final beam splitter, the resulting state will be
\begin{equation}
\begin{split}
\vec{a}_{WW} &= ( \mtx{H} \otimes \mtx{I} ) \mtx{C} ( \mtx{R}_{\phi} \otimes \mtx{I} ) ( \mtx{H} \otimes \mtx{I} ) \vec{a} \\
&= \frac{\alpha}{2} \Bigl( \ket{R,H} + e^{i\phi} \ket{R,V} + \ket{D,H} - e^{i\phi} \ket{D,V} \Bigr) + \frac{\vec{z}'''}{\sqrt{2}} \; .
\end{split}
\end{equation}
The interference pattern is once again lost (i.e., the detection probabilities are independent of $\phi$).  Each of the four modal outcomes occurs now with equal probability, with the likelihood of multiple detections again vanishing as $|\alpha|$ is decreased or $\gamma$ is increased.  Replacing the NOT gate with a unitary gate that is close to the identity will result in a diminished but still discernible interference pattern, so one may consider measuring the path information weakly as well.  So, if there is only partial which-way information, the interference pattern will simply diminish by degrees.

Finally, if one considers the total number of single-detection events on either mode, it is natural to suppose that this should be insensitive to whether the final beam splitter is present or not.  Under the present model, this need not be the case.  Although the total intensity is the same, since the beam splitter constitutes a unitary transformation, the probability of a count on either detector \emph{with} the beam splitter is
\begin{equation}
\begin{split}
P_{\rm MZ} &=1 - \left[ 1 - Q_1\left( |\alpha(1+e^{i\phi})|, 2\gamma \right) \right] \left[ 1 - Q_1\left( |\alpha(1-e^{i\phi})|, 2\gamma \right) \right] \; ,
\end{split}
\end{equation}
while the probability of this event \emph{without} the final beam splitter is
\begin{equation}
\begin{split}
P_{\rm DC} &= 1 - \left[ 1 - Q_1\left( \sqrt{2}|\alpha|, 2\gamma \right) \right] \left[ 1 - Q_1\left( \sqrt{2}|\alpha e^{i\phi}|, 2\gamma \right) \right] \; .
\end{split}
\end{equation}
In general, these two probabilities are different; quantum mechanics predicts that they should be the same.  Nevertheless, for fixed $\gamma$ and $|\alpha| \to 0$ we do find that $P_{\rm MZ}/P_{\rm DC} \to 1$, as one might expect.  This is consistent with previous observation that the approximation $Q_1(2|\alpha|,2\gamma) \approx e^{-2\gamma}(1+4\gamma^2|\alpha|^2)$ is valid when $\gamma|\alpha|$ is small but not necessarily otherwise.  The subtlety of this relationship will elaborated upon further in the next section.


\subsection{General Multimodal States}

The transformation of coherent light via a sequence of linear optical components can be described, in general, by a $d \times d$ unitary matrix $\mtx{U}$.  Without loss of generality, we may suppose that the initial state is of the form
\begin{equation}
\vec{a} = \alpha \begin{bmatrix} 1 \\ 0 \\ \vdots \\ 0 \end{bmatrix} + \frac{1}{\sqrt{2}} \begin{bmatrix} z_1 \\ z_2 \\ \vdots \\ z_d \end{bmatrix} = \alpha \, \vec{\psi} + \frac{\vec{z}}{\sqrt{2}} \; ,
\end{equation}
where $\vec{z}$ is a $d$-dimensional vector of iid standard complex Gaussian random variables and $\alpha \in \mathbb{C}$.  Following the transformation, the new state is
\begin{equation}
\vec{a}' = \alpha \, \mtx{U} \vec{\psi} + \frac{\vec{z}'}{\sqrt{2}} \; ,
\end{equation}
where $\vec{z}' = \mtx{U} \vec{z}$ is, again, a $d$-dimensional vector of iid standard complex Gaussian random variables, owing to the unitarity of $\mtx{U}$.

Detection measurements on the $d$ modes will result in one of $2^d$ possible outcomes.  Let $(n_1, \ldots, n_d) \in \{0,1\}^d$ denote the outcome in which mode 1 has $n_1$ detections, mode 2 has $n_2$, etc., and let $P(n_1, \ldots, n_d)$ denote the probability of this outcome occurring.  Since the random variables $z'_1, \ldots, z'_d$ are statistically independent, this probability is given by
\begin{equation}
P(n_1, \ldots, n_d) = \prod_{i=1}^{d} q_i^{n_i} (1-q_i)^{1-n_i} \; ,
\end{equation}
where $q_i = Q_1(2|\alpha\psi'_i|, 2\gamma)$ is the probability of a threshold crossing event for mode $i$.  (We assume, for simplicity, that all detectors have the same threshold.)

We will be most particularly concerned with single-detection events (i.e., those for which $n_1 + \cdots + n_d = 1$), as these would be interpreted as single-photon detections.  Although such events occur with vanishingly small probability as $|\alpha|$ becomes large (and low probability for $|\alpha|$ small), we may condition, via post-selection, on only such events and thereby obtain a nonvanishing probability.  Specifically, let $p_i$ denote the probability that a single-detection event occurs on mode $i$, given that a single-detection event occurs on any one mode.  It follows that
\begin{equation}
\begin{split}
p_i 
&= q_i \prod_{j \neq i} (1-q_j) \left[ \sum_{k=1}^{d} q_k \prod_{\ell \neq k} (1-q_{\ell}) \right]^{-1}
= \frac{q_i}{1-q_i} \left[ \sum_{k=1}^{d} \frac{q_k}{1-q_k} \right]^{-1} \; ,
\end{split}
\label{eqn:mode_prob}
\end{equation}
provided that $q_k \neq 1$ for all $k$.  Note that, if there exists an $i$ such that $|\psi_i| > |\psi_j|$ for all $j \neq i$, then $p_i \to 1$ as $|\alpha| \to \infty$.  In other words, for bright light only the most probable mode will have a single detection.  For states with no unique maximum, the asymptotic probability is spread uniformly amongst the maxima.  The latter case is consistent with the quantum mechanical predictions, while the former is not.  The right correspondence with quantum mechanics is then to be expected for small or intermediate values of $|\alpha|$.

To examine the validity of our model, we performed linear quantum state tomography (QST) on a random sample of pure states formed by applying unitary matrices drawn from a Haar measure \cite{Mezzadri2007}.  For a given transformed state $\vec{\psi}$, QST was performed using a complete set of $d$-dimensional Hermitian basis matrices $\mtx{B}_1, \ldots, \mtx{B}_{d^2}$ that are orthonormal in the Hilbert-Schmidt inner product.  Each $\mtx{B}_k$ is diagonalized by a unitary matrix $\mtx{U}_k$ such that $(\mtx{U}_k^\dagger \mtx{B}_k \mtx{U}_k)_{ij} = \beta_{ki} \delta_{ij}$.  To measure in this basis, we therefore transformed $\vec{\psi}$ to $\vec{\psi}' = \mtx{U}_k^\dagger \vec{\psi}$ and computed $p_i$ according to equation (\ref{eqn:mode_prob}).  The expectation value of $\mtx{B}_k$ was taken to be $p_1 \beta_{k1} + \cdots + p_d \beta_{k1}$, so the inferred density matrix for $\vec{\psi}$ is defined to be
\begin{equation}
\mtx{\rho} = \sum_{k=1}^{d^2} \mtx{B}_k \, \sum_{i=1}^{d} p_i \, \beta_{ki} \; .
\end{equation}
Since the basis matrices are such that $\mathrm{Tr}[\mtx{B}_1^\dagger \mtx{B}_1] = 1$ and $\mathrm{Tr}[\mtx{B}_k^\dagger \mtx{B}_k] = 0$ for $k > 1$, the trace $\mathrm{Tr}[\mtx{\rho} \mtx{B}_k]$ gives the expectation value of $\mtx{B}_k$, as defined above.  This allows us to identify $\mtx{\rho}$ as playing the role of a quantum mechanical density operator.

\begin{figure}[ht]
\centerline{\scalebox{0.8}{\includegraphics{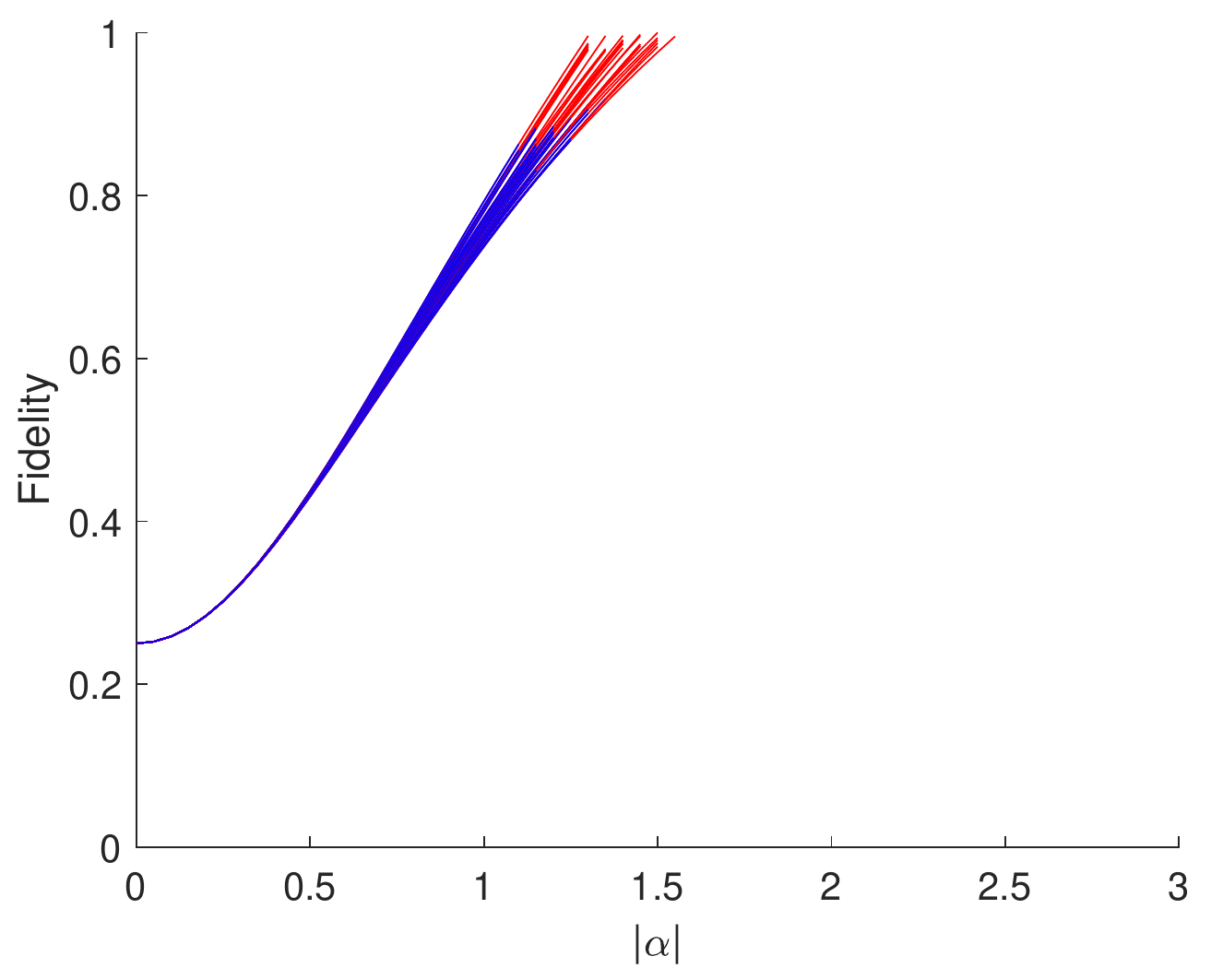}}}
\caption{Plot of the fidelity of the QST solution versus the state amplitude $|\alpha|$ for $\gamma = 1$ over an ensemble of 30 pure states with $d = 4$.  The red portion of the curves indicates where one or more eigenvalues in the density matrix are negative.}
\label{fig:fidelity}
\end{figure}

With $\mtx{\rho}$ so computed, we examined the fidelity, defined by the vector inner product
\begin{equation}
F = \bra{\vec{\psi}} \mtx{\rho} \ket{\vec{\psi}} = \sum_{ij} \psi_i^* \rho_{ij} \psi_j \; ,
\end{equation}
as a function of $d$, $|\alpha|$, and $\gamma$ over an ensemble of pure states $\vec{\psi}$.  In figure \ref{fig:fidelity} we have plotted $F$ versus $|\alpha|$ for $d = 4$ and $\gamma = 1$ for an ensemble of 30 randomly drawn pure states.  For this case, tensor products of the Pauli matrices were used for the orthonormal basis.  We observe that $F  = 1/d$ (corresponding to $p_i = 1/d$) for $|\alpha| = 0$, as expected for pure vacuum noise.  As $|\alpha|$ increases, $F$ increases monotonically to a value near unity.  However, for sufficiently large values of $|\alpha|$ the inferred density matrix acquires negative eigenvalues and becomes invalid.  For general quantum states, taking $|\alpha| \sim \gamma\sqrt{d/2}$ tends to give near unity fidelity, albeit with invalid density matrices.  For ``classical'' states (i.e., those for which $|\psi_i| = 1$ for exactly one index $i$) the density matrix remains valid for all $|\alpha|$ and the fidelity asymptotically approaches unity as $|\alpha| \to \infty$.  Qualitatively similar behavior is found when $\gamma$ is varied while holding $\alpha$ fixed.

Negative eigenvalues in density matrices obtained through linear state tomography are a common occurrence in experimental quantum optics, particularly for low-entropy, high fidelity states.  Their presence might be interpreted as an observed deviation from the Born rule, but they are more commonly ascribed to mere ``experimental inaccuracies and statistical fluctuations'' \cite{James2001}.  According to our model, such results are an inevitable consequence of the parameter regime investigated and the data analysis methods used to infer the quantum state.  Since we compute the probabilities exactly to perform state tomography, we may also conclude that the potential for negative eigenvalues is an intrinsic property of the model and not one due to mere sampling errors.

\begin{figure}[ht]
\centerline{\scalebox{0.8}{\includegraphics{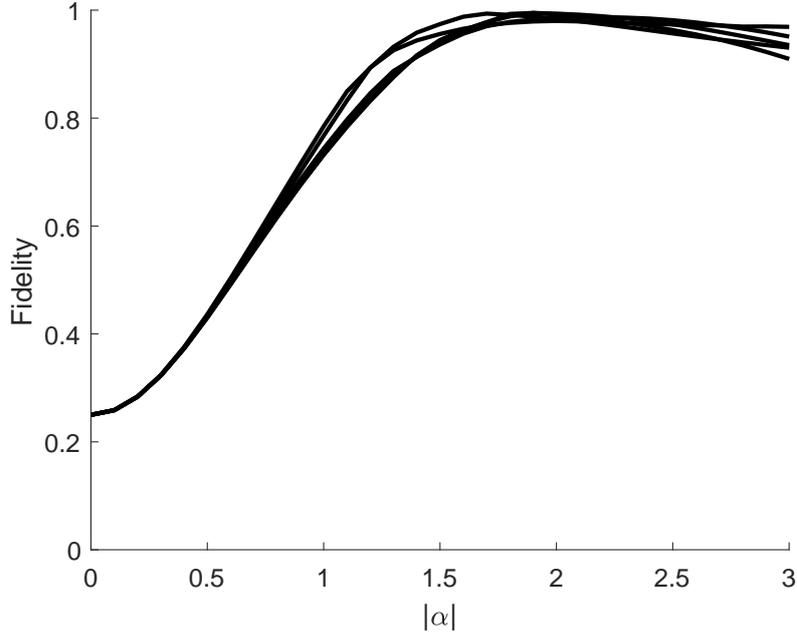}}}
\caption{Plot of the fidelity of the QST solution using the MLE method versus the state amplitude $|\alpha|$ for $\gamma = 1$ over an ensemble of five pure states with $d = 4$.}
\label{fig:fidelityMLE}
\end{figure}

To address the problem of invalid density matrices obtained from linear state tomography, maximum-likelihood estimation (MLE) methods are often used \cite{Hradil1997,Hradil2000,Altepeter2004}.  In this approach, one parameterizes a general, positive semi-definite density matrix and estimates the parameters of this matrix using an optimization scheme based on an assumption of Gaussian errors.  By construction, this approach always yields a valid density matrix.  We reexamined our results using the MLE-based state tomography tools provided by the Kwiat Quantum Information Group \cite{MLEtools}.  The results of five randomly sampled states with $d = 4$ are shown in figure \ref{fig:fidelityMLE}.  For $|\alpha|$ less than or close to unity, the MLE results agree with the previous linear tomography results.  However, for larger values of $|\alpha|$, the curves peak near unity and then slowly decrease as we approach the classical regime of $|\alpha| \gg 1$.  Qualitatively similar behavior is found when $\gamma$ is varied while holding $\alpha$ fixed.  This shows that it is possible to infer density matrix estimates from our model that are both valid and of high fidelity.

The density matrix derived from QST may also be used to examine entanglement.  According to the Peres-Horodecki positive partial transpose (PPT) criterion, a density operator that acts on a tensor product Hilbert space $\mathcal{H}_A \otimes \mathcal{H}_B$ will be separable with respect to $\mathcal{H}_A$ and $\mathcal{H}_B$ if all the eigenvalues of its partial transpose are positive \cite{Peres1996}.  In our case, $\otimes$ is the Kronecker product, $\mathcal{H}_A = \mathbb{C}^{d_A}$, and $\mathcal{H}_B = \mathbb{C}^{d_B}$, for $d_A, d_B \in \mathbb{N}$.  If we write the density matrix $\mtx{\rho}$ as
\begin{equation}
\mtx{\rho} = \sum_{ij} \sum_{k\ell} \rho_{ij;k\ell} \: \vec{e}^A_i (\vec{e}^A_j)^\dagger \otimes \vec{e}^B_k (\vec{e}^B_\ell)^\dagger \; ,
\end{equation}
where $\vec{e}^A_i, \vec{e}^A_j$ and $\vec{e}^B_k, \vec{e}^B_\ell$ are the standard unit vectors in $\mathbb{C}^{d_A}$ and $\mathbb{C}^{d_B}$, respectively, then the partial transpose with respect to $\mathcal{H}_B$ is
\begin{equation}
\mtx{\rho}^{\mathsf{T}_B} = \sum_{ij} \sum_{k\ell} \rho_{ij;k\ell} \: \vec{e}^A_i (\vec{e}^A_j)^\dagger \otimes \vec{e}^B_\ell (\vec{e}^B_k)^\dagger \; .
\end{equation}

\begin{figure}[ht]
\centerline{\scalebox{0.8}{\includegraphics{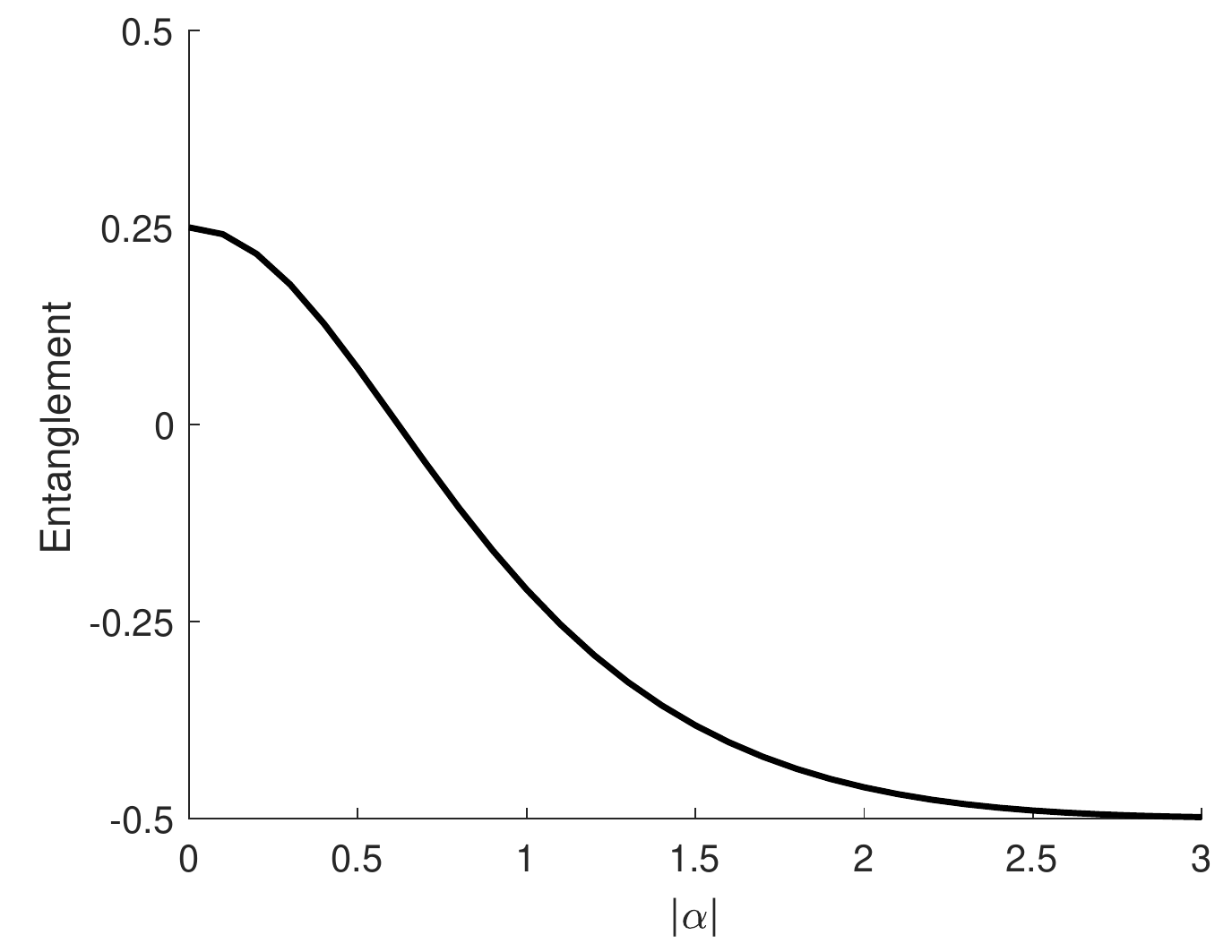}}}
\caption{Plot of the PPT entanglement witness for $d = 4$ versus the state amplitude $|\alpha|$ for a maximally entangled Bell state using a detection threshold of $\gamma = 1$.}
\label{fig:entanglement_witness}
\end{figure}

Negative eigenvalues of the partial transpose are a necessary, though not sufficient, condition for the density matrix to be nonseparable (i.e., entangled).  For certain cases, such as $d_A = d_B = 2$, this condition is also sufficient and therefore may be used as an entanglement witness \cite{Horodecki1996}.  In figure \ref{fig:entanglement_witness} we have plotted the minimum eigenvalue of the partial transpose for $d_A = d_B = 2$ as a function of $|\alpha|$ for a maximally entangled Bell state using a detection threshold of $\gamma = 1$ and the aforementioned MLE method to infer the quantum state.  It is perhaps surprising that, although our inferred density matrix does not have perfect fidelity, it is nevertheless entangled (i.e., nonseparable), as witnessed by the negative eigenvalues of the partial transpose for values of $|\alpha|$ above 0.6.  The behavior for large $|\alpha|$ shows an asymptotic approach to $-0.5$, the value predicted by quantum mechanics for an ideal Bell state.  Qualitatively similar behavior is found when $\gamma$ is varied while holding $\alpha$ fixed.

\begin{figure}[ht]
\centerline{\scalebox{0.8}{\includegraphics{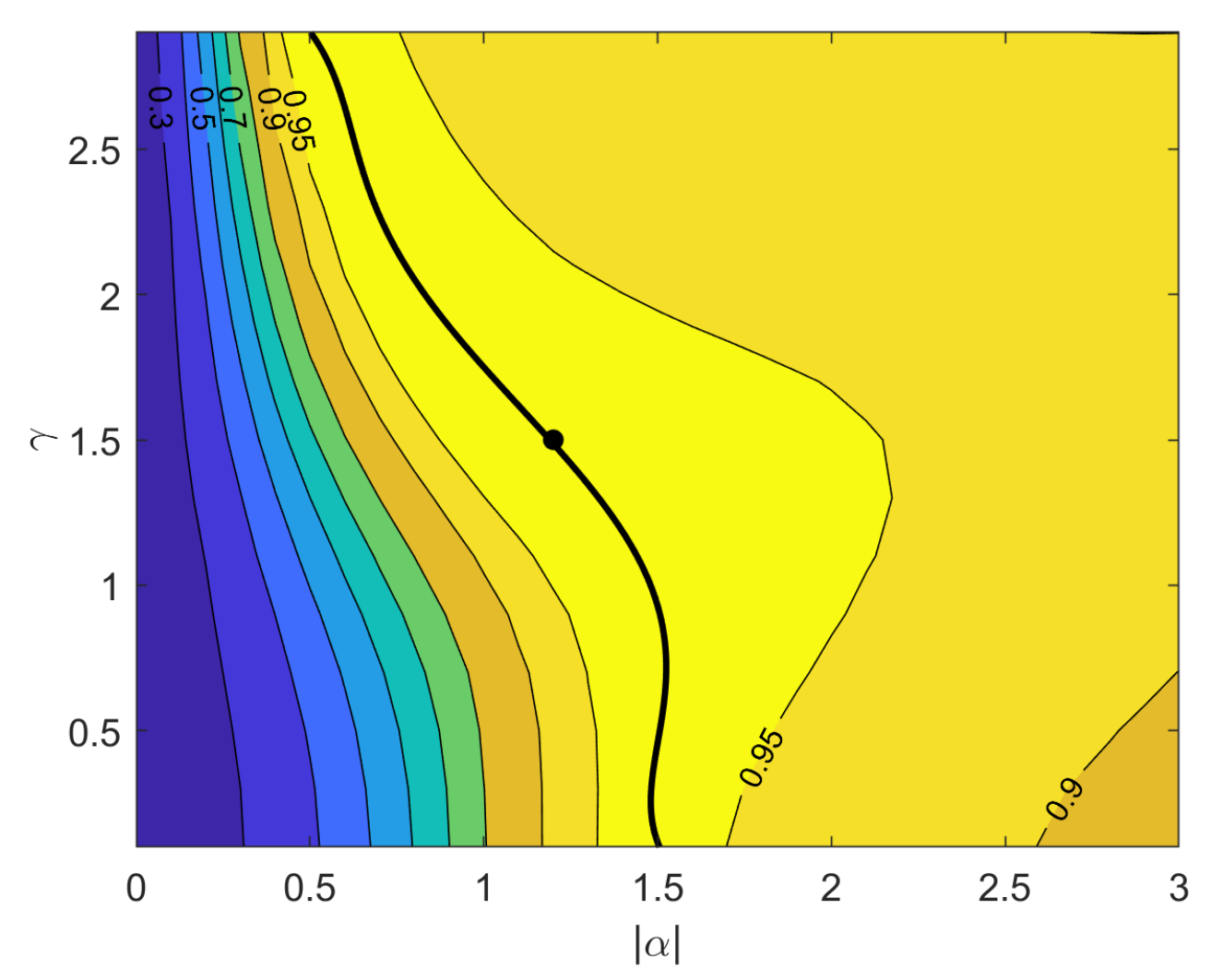}}}
\caption{Contour plot of fidelity using the MLE method versus the state amplitude $|\alpha|$ and detection threshold $\gamma$ over an ensemble of 100 pure states with $d = 4$. The black dot indicates a maximum fidelity of $0.98$ for $|\alpha| = 1.2$ and $\gamma = 1.5$.  The black curve is a spline-interpolation of the local maxima for each given value of $\gamma$.}
\label{fig:fidelityContour}
\end{figure}

In the above examples we have considered fidelity as a function of $|\alpha|$; the dependency on both $|\alpha|$ and $\gamma$ is more subtle.  In figure \ref{fig:fidelityContour} we have plotted the fidelity of the QST solution using the MLE method as a function of both $|\alpha|$ and $\gamma$, averaged over an ensemble of 100 pure states with $d = 4$.  For any given $\gamma > 0$, we see that there is a unique value of $|\alpha|$ giving locally optimal fidelity.  Generally, small values of $|\alpha|$ and large values of $\gamma$ give good, albeit imperfect, fidelity.  The globally optimal fidelity is found to be about 0.98 and occurs near $|\alpha| = 1.2$ and $\gamma = 1.5$.  The fidelity for some states can be higher and qualitatively different.  For example, states with a high degree of symmetry, such as the Bell states, can exhibit fidelities approaching unity when either $|\alpha|$ or $\gamma$ is large.

\begin{figure}[ht]
\centerline{\scalebox{0.8}{\includegraphics{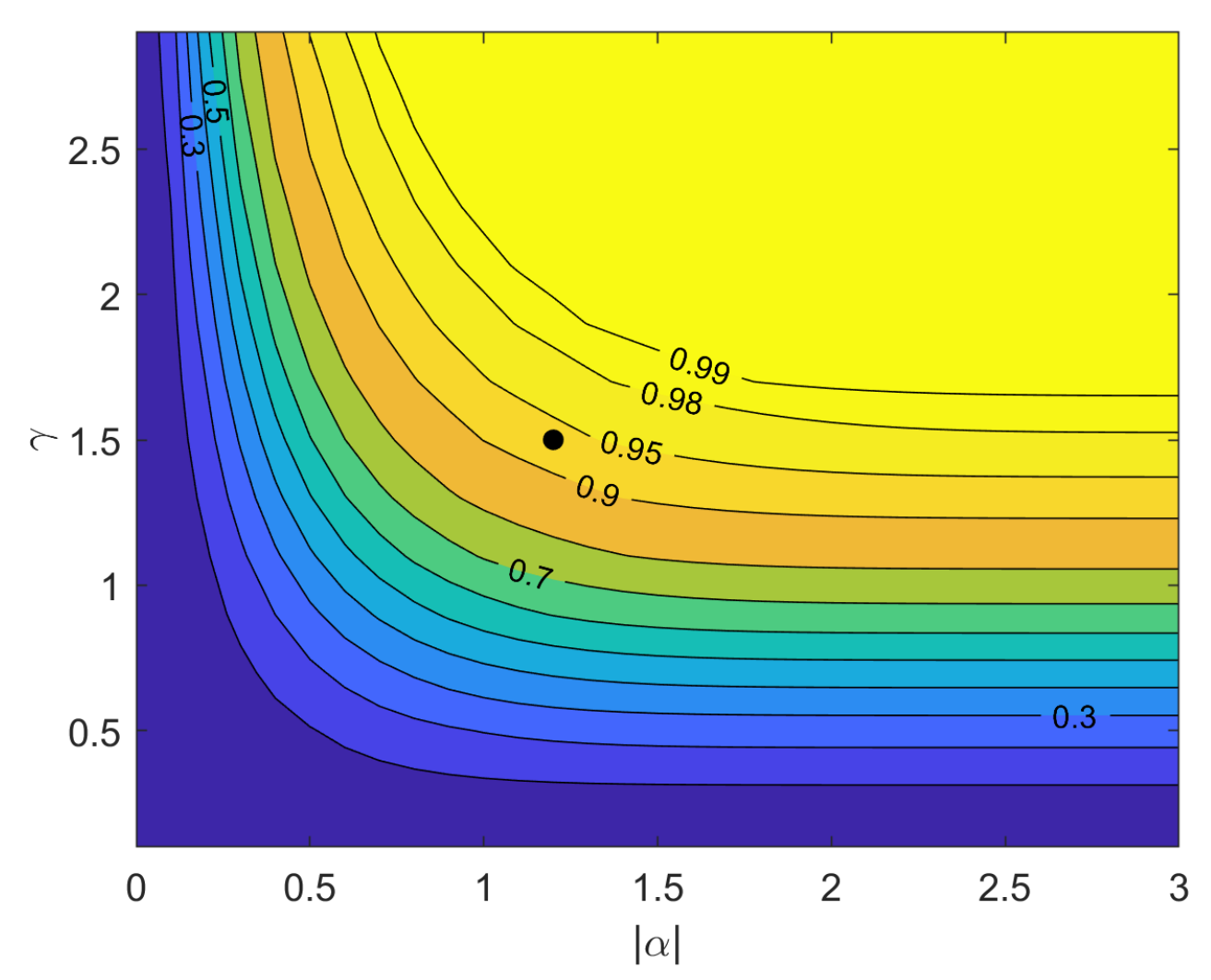}}}
\caption{Plot of visibility versus the state amplitude $|\alpha|$ and detection threshold $\gamma$ for $d = 4$.  The black dot at $|\alpha| = 1.2$ and $\gamma = 1.5$ corresponds to a visibility of 0.94.}
\label{fig:visibilityContour}
\end{figure}

In addition to fidelity, we also considered visibility as a metric of agreement with quantum predictions.  Visibility converges monotonically to one as either $|\alpha|$ or $\gamma$ tends to infinity, but optimal fidelity occurs only for finite values of these parameters, so optimally satisfying both may not be possible.  We have plotted the visibility as a function of $|\alpha|$ and $\gamma$ in figure \ref{fig:visibilityContour}.  The point of maximum fidelity, from figure \ref{fig:fidelityContour}, corresponds to a visibility of 0.94.  Taking $\gamma$ to be large and $|\alpha$, suitably chosen, to be small, can achieve higher visibility while still maintaining a high fidelity, but the two metrics cannot be made arbitrarily close to unity.  Experimentally, fidelities and visibilities as high as 0.99 have been observed, though not necessarily in the same context.  A more detailed analysis of specific experiments would therefore be needed for a proper comparison.

\section{Conclusion}
\label{sec:C}

Assuming a classical zero-point field and deterministic threshold detectors, we have shown that one is able to reproduce many of the experimentally observed phenomena attributed to single photons and thought to be uniquely quantum in nature.  In so doing we have established that such phenomena do, in fact, have classical analogues.  Weak coherent light in combination with a reified zero-point field considered in the single-mode regime are found to give probabilistic outcomes that are in close agreement with the Born rule for single-photon, multi-mode states when post-selection on single-detection events is performed.  This agreement was verified explicitly by performing quantum state tomography and computing the fidelity of the resulting density matrix. The model results are not always in perfect agreement with the idealized quantum mechanical predictions, but they are largely consistent with experimental observations and data analysis methods in the appropriate parameter regimes.  Deviations are, however, expected in regimes in which either the amplitude of the light or the threshold of detection is large.  The best overall agreement appears to be for a set of parameter values in which the amplitude is small and threshold is large. This model therefore provides a local, realistic picture of wave/particle duality and single-photon entanglement that is grounded in a physical and wholly classical model.  A similar classical description of homodyne measurements, temporal behavior, and multi-photon entanglement are left for future work.


\begin{acknowledgments}
This work was supported by the U. S. Office of Naval Research under Grant No.\ N00014-17-1-2107.  The authors also would like to thank the reviewers for their thoughtful comments and suggestions.
\end{acknowledgments}


\bibliographystyle{unsrt}
\bibliography{refs}

\begin{thebibliography}{10}

\bibitem{vonNeumann1931}
J.~von Neumann.
\newblock {\em Mathematische Grundlagen der Quantentheorie}.
\newblock Springer, 1931.

\bibitem{Bell1964}
J.~S. Bell.
\newblock On the {E}instein {P}odolsky {R}osen paradox.
\newblock {\em Physics}, \textbf{1}:195, 1964.

\bibitem{deMuynck2002}
W.~M. de~Muynck.
\newblock {\em Foundations of Quantum Mechanics, An Empiricist Approach}.
\newblock Kluwer Academic Publishers, 2002.

\bibitem{Brunner2014}
N.~Brunner, D.~Cavalcanti, S.~Pironio, V.~Scarani, and S.~Wehner.
\newblock Bell nonlocality.
\newblock {\em Reviews of Modern Physics}, \textbf{2014}:419, 2014.
\newblock
  \href{https://doi.org/10.1103/RevModPhys.86.419}{https://doi.org/10.1103/RevModPhys.86.419}.

\bibitem{Hensen2015}
B.~Hensen et~al.
\newblock Loophole-free bell inequality violation using electron spins
  separated by 1.3 kilometres.
\newblock {\em Nature}, \textbf{526}:682, 2015.
\newblock
  \href{https://doi.org/10.1038/nature15759}{https://doi.org/10.1038/nature15759}.

\bibitem{Giustina2015}
M.~Giustina et~al.
\newblock Significant-loophole-free test of {B}ell's theorem with entangled
  photons.
\newblock {\em Physical Review Letters}, \textbf{115}:250401, 2015.
\newblock
  \href{https://doi.org/10.1103/PhysRevLett.115.250401}{https://doi.org/10.1103/PhysRevLett.115.250401}.

\bibitem{Shalm2015}
L.~K. Shalm et~al.
\newblock Strong loophole-free test of local realism.
\newblock {\em Physical Review Letters}, \textbf{115}:250402, 2015.
\newblock
  \href{https://doi.org/10.1103/PhysRevLett.115.250402}{https://doi.org/10.1103/PhysRevLett.115.250402}.

\bibitem{Rosenfeld2017}
W.~Rosenfeld et~al.
\newblock Event-ready bell test using entangled atoms simultaneously closing
  detection and locality loopholes.
\newblock {\em Physical Review Letters}, \textbf{119}:010402, 2017.
\newblock
  \href{https://doi.org/10.1103/PhysRevLett.119.010402}{https://doi.org/10.1103/PhysRevLett.119.010402}.

\bibitem{Bierhorst2018}
P.~Bierhorst et~al.
\newblock Experimentally generated randomness certified by the impossibility of
  superluminal signals.
\newblock {\em Nature}, \textbf{556}:223, 2018.
\newblock
  \href{https://doi.org/10.1038/s41586-018-0019-0}{https://doi.org/10.1038/s41586-018-0019-0}.

\bibitem{Landau&Lifshitz4}
V.~B. Berestetskii, L.~P. Pitaevskii, and E.~M. Lifshitz.
\newblock {\em Quantum Electrodynamics}, volume~4.
\newblock Elsevier, 2nd edition, 1982.

\bibitem{Milonni1994}
P.~W. Milonni.
\newblock {\em The Quantum Vacuum. An Introduction to Quantum Electrodynamics}.
\newblock Academic Press, 1994.

\bibitem{Landsman2008}
N.~P. Landsman.
\newblock {\em Compendium of Quantum Physics}, chapter The Born rule and its
  interpretation.
\newblock Springer, 2008.

\bibitem{Born1926}
M.~Born.
\newblock Zur quantenmechanik der sto{\ss}vorg{\"{a}}nge.
\newblock {\em Zeitschrift f{\"{u}}r Physik}, 37:863, 1926.
\newblock
  \href{https://doi.org/10.1007/BF01397477}{https://doi.org/10.1007/BF01397477}.

\bibitem{Gleason1957}
A.~M. Gleason.
\newblock Measures on the closed subspaces of a {H}ilbert space.
\newblock {\em Journal of Mathematical Mechanics}, \textbf{6}:885, 1957.
\newblock
  \href{https://doi.org/10.1007/978-94-010-1795-4_7}{https://doi.org/10.1007/978-94-010-1795-4\_7}.

\bibitem{Deutsch1999}
D.~Deutsch.
\newblock Quantum theory of probability and decisions.
\newblock {\em Proceedings of the Royal Society A}, 455:3129, 1999.
\newblock
  \href{https://doi.org/10.1098/rspa.1999.0443}{https://doi.org/10.1098/rspa.1999.0443}.

\bibitem{Barnum2000}
H.~Barnum, C.~M. Caves, J.~Finkelstein, C.~A. Fuchs, and R.~Schack.
\newblock Quantum probability from decision theory?
\newblock {\em Proceedings of The Royal Society}, \textbf{A456}:1175, 2000.
\newblock
  \href{https://doi.org/10.1098/rspa.2000.0557}{https://doi.org/10.1098/rspa.2000.0557}.

\bibitem{Zurek2005}
W.~H. Zurek.
\newblock Probabilities from entanglement, {B}orn's rule $p_k = |\psi_k| ^2$
  from envariance.
\newblock {\em Physical Review A}, \textbf{71}:052105, 2005.
\newblock
  \href{https://doi.org/10.1103/PhysRevA.71.052105}{https://doi.org/10.1103/PhysRevA.71.052105}.

\bibitem{Schlosshauer2005}
M.~Schlosshauer and A.~Fine.
\newblock On {Z}urek's derivation of the {B}orn rule.
\newblock {\em Foundations of Physics}, \textbf{35}:197, 2005.
\newblock
  \href{https://doi.org/10.1007/s10701-004-1941-6}{https://doi.org/10.1007/s10701-004-1941-6}.

\bibitem{Masanes2019}
L.~Masanes, T.~D. Galley, and M.~P. M{\"{u}}ller.
\newblock The measurement postulates of quantum mechanics are operationally
  redundant.
\newblock {\em Nature Communications}, 10:1361, 2019.
\newblock
  \href{https://doi.org/10.1038/s41467-019-09348-x}{https://doi.org/10.1038/s41467-019-09348-x}.

\bibitem{Allahverdyan2013}
A.~E. Allahverdyan, R.~Balian, and T.~M. Nieuwenhuizen.
\newblock Understanding quantum measurement from the solution of dynamical
  models.
\newblock {\em Physics Reports}, \textbf{525}:1, 2013.
\newblock
  \href{https://doi.org/10.1016/j.physrep.2012.11.001}{https://doi.org/10.1016/j.physrep.2012.11.001}.

\bibitem{TheDice}
L.~de~la Pe\~{n}a and A.~M. Cetto.
\newblock {\em The Quantum Dice: An Introduction to Stochastic
  Electrodynamics}.
\newblock Kluwer, 1995.

\bibitem{Casado1997}
A.~Casado, T.~W. Marshall, and E.~Santos.
\newblock Parametric downconversion experiments in the {W}igner representation.
\newblock {\em Journal of the Optical Society of America B}, \textbf{14}:494,
  1997.
\newblock
  \href{https://doi.org/10.1364/JOSAB.14.000494}{https://doi.org/10.1364/JOSAB.14.000494}.

\bibitem{Marshall1988}
T.~W. Marshall and E.~Santos.
\newblock Stochastic optics: A reaffirmation of the wave nature of light.
\newblock {\em Foundations of Physics}, \textbf{18}:185, 1988.
\newblock
  \href{https://doi.org/10.1007/BF01882931}{https://doi.org/10.1007/BF01882931}.

\bibitem{Adenier2009}
G.~Adenier.
\newblock Violation of {B}ell inequalities as a violation of fair sampling in
  threshold detectors.
\newblock In {\em AIP Conference Proceedings \textbf{1101}}, page~8, 2009.
\newblock
  \href{https://doi.org/10.1063/1.3109977}{https://doi.org/10.1063/1.3109977}.

\bibitem{LaCour2014}
B.~{La Cour}.
\newblock A locally deterministic, detector-based model of quantum measurement.
\newblock {\em Foundations of Physics}, \textbf{44}:1059, 2014.
\newblock
  \href{https://doi.org/10.1007/s10701-014-9829-6}{https://doi.org/10.1007/s10701-014-9829-6}.

\bibitem{Khrennikov}
A.~Khrennikov.
\newblock {\em Beyond Quantum}.
\newblock Pan Stanford Publishing, 2014.

\bibitem{Planck1911}
M.~Planck.
\newblock Eine neue strahlungshypothese.
\newblock {\em Verhandlungen der Deutschen Physikalischen Gesellschaft},
  13:138, 1911.

\bibitem{Marshall1963}
T.~W. Marshall.
\newblock Random electrodynamics.
\newblock {\em Proceedings of the Royal Society}, A276:475, 1963.
\newblock
  \href{http://doi.org/10.1098/rspa.1963.0220}{http://doi.org/10.1098/rspa.1963.0220}.

\bibitem{Ibison1996}
M.~Ibison and B.~Haisch.
\newblock Quantum and classical statistics of the electromagnetic zero-point
  field.
\newblock {\em Physical Review A}, \textbf{54}:2737, 1996.
\newblock
  \href{https://doi.org/10.1103/PhysRevA.54.2737}{https://doi.org/10.1103/PhysRevA.54.2737}.

\bibitem{Lasota&Mackey}
A.~Lasota and M.~Mackey.
\newblock {\em Chaos, Fractals, and Noise: Stochastic Aspects of Dynamics}.
\newblock Springer, 2nd edition, 1998.

\bibitem{Franca&Marshall1988}
H.~M. Fran{\c{c}}a and T.~W. Marshall.
\newblock Excited states in stochastic electrodynamics.
\newblock {\em Physical Review A}, 38:3258, 1988.
\newblock
  \href{https://doi.org/10.1103/PhysRevA.38.3258}{https://doi.org/10.1103/PhysRevA.38.3258}.

\bibitem{Ossiander2018}
M.~Ossiander et~al.
\newblock Absolute timing of the photoelectric effect.
\newblock {\em Nature}, 561:374, 2018.
\newblock
  \href{https://doi.org/10.1038/s41586-018-0503-6}{https://doi.org/10.1038/s41586-018-0503-6}.

\bibitem{Shibata2015}
H.~Shibata, K.~Shimizu, H.~Takesue, and Y.~Tokura.
\newblock Ultimate low system dark count rate for superconducting nanowire
  single-photon detector.
\newblock {\em Optics Letters}, 40:3428, 2015.
\newblock
  \href{https://doi.org/10.1364/OL.40.003428}{https://doi.org/10.1364/OL.40.003428}.

\bibitem{Johnson1994}
N.~L. Johnson, S.~Kotz, and N.~Balakrishnan.
\newblock {\em Continuous Univariate Distributions}.
\newblock John Wiley and Sons, 1994.

\bibitem{Cahill1969}
K.~Cahill and R.~Glauber.
\newblock Density operators and quasiprobability distributions.
\newblock {\em Physical Review}, \textbf{177}:1882, 1969.
\newblock
  \href{https://doi.org/10.1103/PhysRev.177.1882}{https://doi.org/10.1103/PhysRev.177.1882}.

\bibitem{Fujii2004}
K.~Fujii and T.~Suzuki.
\newblock A new symmetric expression of {W}eyl ordering.
\newblock {\em Modern Physics Letters A}, \textbf{19}:827, 2004.
\newblock
  \href{https://doi.org/10.1142/S021773230401374X}{https://doi.org/10.1142/S021773230401374X}.

\bibitem{Campbell1984}
B.~F. Levine, D.~G. Bethea, and J.~C. Campbell.
\newblock Near room temperature 1.3um single photon counting with a ingaas
  avalanche photodiode.
\newblock {\em Electronics Letters}, \textbf{20}:596, 1984.
\newblock
  \href{https://doi.org/10.1049/el:19840411}{https://doi.org/10.1049/el:19840411}.

\bibitem{Oh2010}
J.~Oh, C.~Anonelli, M.~Tur, and M.~Brodsky.
\newblock Method for characterizing single photon detectors in saturation
  regime by cw laser.
\newblock {\em Optics Express}, \textbf{18}:5906, 2010.
\newblock
  \href{https://doi.org/10.1364/OE.18.005906}{https://doi.org/10.1364/OE.18.005906}.

\bibitem{Loudon1980}
R.~Loudon.
\newblock Non-classical effects in the statistical properties of light.
\newblock {\em Reports on Progress in Physics}, \textbf{43}:913, 1980.
\newblock
  \href{https://doi.org/10.1088/0034-4885/43/7/002}{https://doi.org/10.1088/0034-4885/43/7/002}.

\bibitem{Grangier2019}
R.~W. Boyd, S.~G. Lukishova, and V.~N. Zadkov, editors.
\newblock {\em The First Single Photon Sources and Single Photon Interference
  Experiments}, chapter~1.
\newblock Springer, 2019.

\bibitem{Grangier1986}
P.~Grangier, G.~Roger, and A.~Aspect.
\newblock Experimental evidence for a photon anticorrelation effect on a beam
  splitter: {A} new light on single-photon interferences.
\newblock {\em Europhysics Letters}, \textbf{1}:173, 1986.
\newblock
  \href{https://doi.org/10.1209/0295-5075/1/4/004}{https://doi.org/10.1209/0295-5075/1/4/004}.

\bibitem{Thorn2004}
J.~J. Thorn et~al.
\newblock Observing the quantum behavior of light in an undergraduate
  laboratory.
\newblock {\em American Journal of Physics}, \textbf{79}:1210, 2004.
\newblock
  \href{https://doi.org/10.1119/1.1737397}{https://doi.org/10.1119/1.1737397}.

\bibitem{Kwiat1998}
P.~Kwiat and H.~Weinfurter.
\newblock Embedded {B}ell-state analysis.
\newblock {\em Physical Review A}, \textbf{58}:R2623, 1998.
\newblock
  \href{https://doi.org/10.1103/PhysRevA.58.R2623}{https://doi.org/10.1103/PhysRevA.58.R2623}.

\bibitem{Jacques2007}
V.~Jacques, E.~Wu, F.~Grosshans, F.~Treussart, P.~Grangier, A.~Aspect, and
  J.-F. Roch.
\newblock Experimental realization of {W}heeler's delayed-choice gedanken
  experiment.
\newblock {\em Science}, \textbf{315}:966, 2007.
\newblock
  \href{https://doi.org/10.1126/science.1136303}{https://doi.org/10.1126/science.1136303}.

\bibitem{Mezzadri2007}
F.~Mezzadri.
\newblock How to generate random matrices from the classical compact groups.
\newblock {\em Notices of the AMS}, 54:592, 2007.
\newblock
  \href{http://www.ams.org/notices/200705/index.html}{http://www.ams.org/notices/200705/index.html}.

\bibitem{James2001}
D.~F.~V. James, P.~G. Kwiat, W.~J. Munro, and A.~G. White.
\newblock Measurement of qubits.
\newblock {\em Physical Review A}, \textbf{64}:052312, 2001.
\newblock
  \href{https://doi.org/10.1103/PhysRevA.64.052312}{https://doi.org/10.1103/PhysRevA.64.052312}.

\bibitem{Hradil1997}
Z.~Hradil.
\newblock Quantum-state estimation.
\newblock {\em Physical Review A}, \textbf{55}:R1561(R), 1997.
\newblock
  \href{https://doi.org/10.1103/PhysRevA.55.R1561}{https://doi.org/10.1103/PhysRevA.55.R1561}.

\bibitem{Hradil2000}
Z.~Hradil, J.~Summhammer, G.~Badurek, and H.~Rauch.
\newblock Reconstruction of the spin state.
\newblock {\em Physical Review A}, \textbf{62}:014101, 2000.
\newblock
  \href{https://doi.org/10.1103/PhysRevA.62.014101}{https://doi.org/10.1103/PhysRevA.62.014101}.

\bibitem{Altepeter2004}
J.~B. Altepeter, D.~F.~V. James, and P.~G. Kwiat.
\newblock {\em Lecture Notes in Physics}, chapter Quantum State Estimation.
\newblock Springer, Berlin, 2004.

\bibitem{MLEtools}
{Kwiat Quantum Information Group}.
\newblock \textit{Guide to Quantum State Tomography}.
\newblock \url{http://research.physics.illinois.edu/QI/Photonics/Tomography/}
  {Accessed 20 December 2019}.

\bibitem{Peres1996}
A.~Peres.
\newblock Separability criterion for density matrices.
\newblock {\em Physical Review Letters}, 77:1413, 1996.
\newblock
  \href{https://doi.org/10.1103/PhysRevLett.77.1413}{https://doi.org/10.1103/PhysRevLett.77.1413}.

\bibitem{Horodecki1996}
P.~Horodecki M.~Horodecki and R.~Horodecki.
\newblock Separability of mixed states: necessary and sufficient conditions.
\newblock {\em Physics Letters A}, 223:1, 1996.
\newblock
  \href{https://doi.org/10.1016/S0375-9601(96)00706-2}{https://doi.org/10.1016/S0375-9601(96)00706-2}.

\end{thebibliography}

\end{document}